\documentclass[journal]{IEEEtran}
\ifCLASSINFOpdf
\else
\fi

\usepackage{booktabs} % For formal tables
\usepackage{subfig}
\usepackage{graphicx}
\usepackage{algorithm}
\usepackage{algpseudocode}% 
\usepackage{xcolor}
\usepackage{amsmath}
\newcommand{\etal}{\textit{et al}. }
\usepackage{mathtools}
\usepackage{multirow}
\usepackage{graphics}
\usepackage{enumitem}
\usepackage[utf8]{inputenc}
\usepackage{ragged2e}
\usepackage{amsmath,amssymb}
\DeclareMathOperator{\E}{\mathbb{E}}
\usepackage{xcolor,cite,etoolbox}
\definecolor{mypurple}{rgb}{0.4392, 0.1882, 0.6275}
\definecolor{mygreen}{rgb}{0, 0.6902, 0.3137}
\makeatletter 
\pretocmd\@bibitem{\color{black}\csname keycolor#1\endcsname}{}{\fail}
\newcommand\citecolor[1]{\@namedef{keycolor#1}{\color{black}}}
\makeatother
\begin{document}
\title{EEGFuseNet: Hybrid Unsupervised Deep Feature Characterization and Fusion for High-Dimensional EEG with An Application to Emotion Recognition}
\author{\IEEEauthorblockN{Zhen Liang\textsuperscript{1,2,3,}\IEEEauthorrefmark{1},
Rushuang Zhou\textsuperscript{1,2,}\IEEEauthorrefmark{2},
Li Zhang\textsuperscript{1,2,}\IEEEauthorrefmark{3},
Linling Li\textsuperscript{1,2,}\IEEEauthorrefmark{4},\\
Gan Huang\textsuperscript{1,2,}\IEEEauthorrefmark{5}, 
Zhiguo Zhang\textsuperscript{1,2,4,5,}\IEEEauthorrefmark{6} and
Shin Ishii\textsuperscript{3,6,}\IEEEauthorrefmark{7}}\\
\medskip
\IEEEauthorblockA{\small{{\textsuperscript{1}School of Biomedical Engineering, Health Science Center, Shenzhen University, Shenzhen, Guangdong 518060, China\\
\textsuperscript{2}Guangdong Provincial Key Laboratory of Biomedical Measurements and Ultrasound Imaging, Shenzhen, Guangdong 518060, China\\
\textsuperscript{3}Graduate School of Informatics, Kyoto University, Kyoto 606-8501, Japan\\
\textsuperscript{4}Marshall Laboratory of Biomedical Engineering, Shenzhen, Guangdong 518060, China\\
\textsuperscript{5}Peng Cheng Laboratory, Shenzhen, Guangdong 518055, China\\
\textsuperscript{6} ATR Neural Information Analysis Laboratories, Kyoto 619-0288, Japan\\
\medskip
Email: \IEEEauthorrefmark{1}janezliang@szu.edu.cn,
\IEEEauthorrefmark{2}2018222087@szu.edu.cn,
\IEEEauthorrefmark{3}lzhang@szu.edu.cn,\\
\IEEEauthorrefmark{4}lilinling@szu.edu.cn,
\IEEEauthorrefmark{5}huanggan@szu.edu.cn,
\IEEEauthorrefmark{6}zgzhang@szu.edu.cn,
\IEEEauthorrefmark{7}ishii@i.kyoto-u.ac.jp}}}}

\maketitle

\begin{abstract}
How to effectively and efficiently extract valid and reliable features from high-dimensional electroencephalography (EEG), particularly how to fuse the spatial and temporal dynamic brain information into a better feature representation, is a critical issue in brain data analysis. \textcolor{black}{Most current EEG studies work in a task driven manner and explore the valid EEG features with a supervised model, which would be limited by the given labels to a great extent.} In this paper, we propose a practical hybrid unsupervised deep \textcolor{black}{convolutional recurrent generative adversarial network} based EEG feature characterization and fusion model, which is termed as EEGFuseNet. EEGFuseNet is trained in an unsupervised manner, and deep EEG features covering \textcolor{black}{both} spatial and temporal dynamics are automatically characterized. Comparing to the \textcolor{black}{existing} features, the \textcolor{black}{characterized} deep EEG features could be considered to be more generic and independent of any specific EEG task. The performance of the extracted deep and low-dimensional features by EEGFuseNet is carefully evaluated in an unsupervised emotion recognition application based on three public emotion databases. The results demonstrate the proposed EEGFuseNet is a robust and reliable model, which is easy to train and performs efficiently in the representation and fusion of dynamic EEG features. In particular, EEGFuseNet is established as an optimal unsupervised fusion model with promising cross-subject emotion recognition performance. It proves EEGFuseNet is capable of characterizing and fusing deep features that imply comparative cortical dynamic significance corresponding to the changing of different emotion states, and also demonstrates the possibility of realizing EEG based cross-subject emotion recognition in a pure unsupervised manner.

\end{abstract}

\begin{IEEEkeywords}
Electroencephalography; Information Fusion; Hybrid Deep Encoder-Decoder Network; CNN-RNN-GAN; Unsupervised; Emotion Recognition.
\end{IEEEkeywords}

% \fi
\IEEEpeerreviewmaketitle

\section{Introduction}
\IEEEPARstart{E}{lectroencephalography} (EEG) is a vital measurement of brain activity that could reflect the activities of neuron dynamics originated from the central nervous system and respond rapidly to different brain states \cite{alarcao2017emotions}. Recently, EEG based emotion recognition has become an increasingly important topic for human emotion understanding, regulation, and management \cite{hu2020video}. However, due to the microvolt-range amplitude of EEG, the collected EEG data is easily contaminated with noises (e.g., physiological artifacts or non-physiological artifacts). In spite of the large number of studies working on EEG-based emotion decoding, how to effectively and efficiently extract valid and useful EEG features from the collected data is still a big challenging. For example, how to fuse the EEG signals collected at different brain locations and at different time points in an efficient approach remains unclear. In general, there are three types of information fusion in the processing of EEG signals. \textbf{(1) Fusion of spatial information:} based on the given time point(s), fuse the relationship dynamics (e.g. correlation) of the EEG signals at different brain regions. The commonly used methods \textcolor{black}{include} connectivity \cite{haufe2013critical}, microstate \cite{milz2016functional}, or other topographic analysis \cite{castelnovo2016scalp,ma2017eeg}, that interpret cortical region communication behavior by assessing the interaction functions between cortical areas and measuring the direction and strength of the interactions. \textcolor{black}{\textbf{(2) Fusion of temporal information:} at the specific brain regions or EEG electrodes, truncate continuous time-series EEG signals into short data segments and fuse the EEG samples at different time points by calculating time-domain features (such as statistical patterns and shape information), frequency-domain features (such as power spectral features), or time-frequency features. Two well-known examples are short-time Fourier transform (STFT) \cite{ramos2020feature} and wavelet analysis \cite{islam2019wavelet}, both of which compute time-frequency dynamics of the data by performing successive calculations and measuring the data interaction along the time. This type of fusion applies to commonly used EEG features in the time-, frequency-, or time-frequency domain at one specific electrode. \textbf{(3) Fusion of both spatial and temporal information:} not only assess cortical region interaction but also estimate dynamic cortical involvement in a serial reaction time. Currently, a number of studies try to characterize both temporal and spatial information in a sequential approach, in which the temporal information is characterized from EEG data at each electrode in the first step and the spatial information is characterized in the second step by measuring the relationship between any two electrodes or among a set of electrodes in terms of the characterized temporal information at each EEG electrode in the first step. Although it is possible to estimate spatial features and temporal features, such an approach would be limited by the pre-defined sequential relationship between the spatial and temporal information and fail to effectively extract and fuse useful but latent information from a joint temporal-spatial domain. Thus, it is still a substantial challenge in current feature extraction methods for EEG signals considering the factors of validity and reliability, which needs to be tackled urgently.}

\textcolor{black}{Deep learning in neural networks provides a good solution to characterize and fuse deep semantic features from the input data and has achieved tremendous success in solving EEG based emotion decoding problems \cite{jirayucharoensak2014eeg, zheng2015investigating, schirrmeister2017deep,song2018eeg,cimtay2020investigating}. For example, Jirayucharoensak \etal \cite{jirayucharoensak2014eeg} introduced a deep learning network with a stack of three autoencoders and two softmax classifiers to perform EEG-based emotion classification, where an improvement of a three-level emotion classification (arousal: 46.03\%; valence: 49.52\%) was demonstrated under a comparison with two baseline methods (support vector machine (SVM) and naïve Bayes). Zheng and Lu \cite{zheng2015investigating} constructed deep belief networks (DBNs) to investigate the critical frequency bands and channels in EEG signals and select the optimal ones by considering the weight distribution learnt from the trained DBNs. Song \etal \cite{song2018eeg} presented a novel dynamic graph convolutional neural network (DGCNN) to solve a multichannel EEG emotion recognition problem, where the discriminant EEG features as well as the intrinsic relationship were learnt. This model manifested that a non-linear deep neural network (DNN) is a powerful tool in solving EEG signals which are highly non-linear in nature. Cimtay and Ekmekcioglu \cite{cimtay2020investigating} adopted a pretrained state-of-the-art CNN, InceptionResnetV2, to extract useful and hidden features from the raw EEG signals. In these studies, the effectiveness of deep feature extraction and representation on EEG signals has been well demonstrated.} However, all the above-mentioned studies were supervised learning based, where a great size of training samples with emotion labels was highly required, especially for deep networks. For example, a smaller size of training samples would make the deep network fail to generalize well due to the overfitting problem. Not only the network design but also the sample size would significantly affect the network performance. Unlike multimedia sources which can be easily obtained from social media platforms like YouTube, it is unrealistic to collect a huge number of EEG signals from different participants and manually annotate each sample with emotional labels in the real-world application scenarios. Also, there is a risk to induce “label noise” during the sample annotation process \cite{luo2017unsupervised}. Unsupervised learning would provide a more natural approach to decode EEG signals and is more aligned with human learning mechanism that requires useful information from the available samples without any associated teachers \cite{barlow1989unsupervised}. \textcolor{black}{How to appropriately characterize EEG signals is one of the most important part in an unsupervised EEG decoding model, which should be able to explore an optimal feature set and achieve a good unsupervised learning performance even in the absence of label guidance.} An improper feature representation would lead to a wrong estimation of relationship structure among the samples. Recently, Liang \etal \cite{liang2019unsupervised} introduced a novel hypergraph-based unsupervised EEG decoding model for human emotion recognition \textcolor{black}{using traditional and shallow EEG features such as statistical features, Hjorth features, frequency bandpowers, energy and entropy properties.} Unfortunately, traditional and shallow features mostly rely on heuristics, prior knowledge and experience, and the modelling performance would be limited. Also, traditional features may fail to efficiently elicit the complicated and non-linear patterns from the raw EEG data. 

\textcolor{black}{There is now a need for valid and reliable deep feature extraction method for time-series high-dimensional EEG signals (a number of electrodes placed along the scalp $\times$ sampling points at a high sampling rate) under an information fusion of spatial and temporal cortical dynamics, and meet this need in an unsupervised manner.} Emerging progress in unsupervised based encoder-decoder networks (the basic ideas of a deep encoder-decoder architecture are introduced in Appendix A of Supplementary Materials\footnote[1]{https://drive.google.com/file/d/1pIphu7LD5MrHsN6GRy1-xZcc9CtV24tm/view?usp=sharing}) has offered a huge success in feature characterization and representation for images \cite{tao2015unsupervised}, videos \cite{kiran2018overview}, and audios \cite{deng2014autoencoder}. The encoder-decoder structures perform excellently in the aspects of highly non-linear feature extraction. Wen and Zhang \cite{wen2018deep} proposed a deep autoencoder based DNN to learn low-dimensional features from high-dimensional EEG data in an unsupervised manner and adopted several commonly used supervised classifiers to demonstrate an improvement of the detection accuracy could be achieved. Similarly, considering the non-stationary and chaotic behavior of the high-dimensional EEG signals, Shoeibi \etal \cite{shoeibi2021comprehensive} developed a convolutional autoencoder for EEG feature learning and showed an accurate and reliable performance in a computer-aided diagnosis system. Instead of directly using EEG raw signals, Tabar and Halici \cite{tabar2016novel} converted high-dimensional EEG data to two-dimensional images by STFT and fed into a stack autoencoder network to solve a classification problem. \textcolor{black}{In this study, we propose a novel unsupervised EEG feature extraction method (termed as EEGFuseNet below) and solve the emotion recognition problem using hypergraph theory. The proposed EEGFuseNet based hypergraph decoding framework includes two parts. \textbf{(1) EEGFuseNet}. An efficient hybrid deep encoder-decoder network architecture is proposed to characterize non-stationary time-series EEG signals. The joint-information of spatial and temporal dynamics is fused in an effective manner and the useful but latent spatial-temporal dynamic information are characterized. The proposed hybrid network incorporates different sources of feature information through integrating CNN, recurrent neural network (RNN) and generative adversarial network (GAN) in a smart hybrid manner. Specifically, based on the features extracted by CNN from raw EEG signals, RNN is adopted to enhance the feature representation by exploring the potential feature relationships at temporal adjacencies. To improve the training performance, GAN is incorporated to improve the training process of the CNN-RNN network through dynamic updates in an unsupervised manner, which is potentially beneficial to high-quality feature generation. The extracted deep features could represent the spatial relationship among the channels and the dependencies of the signals collected at adjacent time points. \textbf{(2) Hypergraph decoding model}. An effective hypergraph decoding model is developed to classify emotions based on the characterized deep features, where the complex relations of brain dynamics under various emotion statuses are measured and the EEG-based emotion classification problem is solved. Specifically, we measure the sample relationships in terms of the characterized deep features in the hypergraph construction, where the EEG samples that share similar properties are connected by hyperedges (the hyperedges are more flexible to describe group relationships). Following the hypergraph partitioning rule, the hypergraph Laplacian is then computed and optimized, where the connections among the hyperedges that share similar properties are grouped into one cluster while the connections among the hyperedges that share different properties are grouped into different clusters.}

\textcolor{black}{We evaluate the performance of the proposed EEGFuseNet based hypergraph decoding framework with an emotion recognition application on three well-known public databases and compare to the other state-of-the-art methods. The results show the generalizability of the proposed unsupervised framework is established and the individual difference is well solved in the cross-subject task. Regardless of the existing deep unsupervised EEG networks, to our knowledge, there is no example of studies where a solid and thorough exploration on hybrid deep configuration for converting high-dimensional EEG signals to low-dimensional valid and reliable feature representation in a fusion and unsupervised manner has been conducted. The proposed EEGFuseNet together with hypergraph decoding would be beneficial to brain decoding applications and offer a pure unsupervised framework for EEG feature extraction, fusion and classification for other use-cases.} \textcolor{black}{The major novelties of this work are as follows. (1) A hybrid unsupervised deep EEGFuseNet is proposed, which serves as a fundamental framework for high-dimensional EEG feature characterization and fusion. A valid and reliable deep EEG feature representation is formed to cover both spatial and temporal dynamics in brain activities, under a consideration of cortical region interactions and cortical involvement in a serial reaction time. (2) A unified unsupervised EEGFuseNet based hypergraph decoding framework is established, and its feasibility and effectiveness in solving brain decoding applications are demonstrated.} \textcolor{black}{(3) A cross-individual task of EEG-based emotion recognition is employed to validate the generality of the proposed unified unsupervised framework on three famous affective databases for the individual difference problem which is common in brain studies. On all three databases, the proposed method outperforms the existing unsupervised methods and achieves a comparable performance comparing to the existing supervised methods without transfer learning strategy.}

% \begin{itemize}
% \color{black}
%   \item A hybrid unsupervised deep EEGFuseNet is proposed, which could be used as a fundamental framework for high-dimensional EEG feature characterization and fusion study. A valid and reliable deep EEG feature representation is formed to cover both spatial and temporal dynamics in brain activities, under a consideration of cortical region interactions and cortical involvement in a serial reaction time.
%   \item A unified unsupervised EEGFuseNet based hypergraph decoding framework is established, and the feasibility in solving brain decoding applications is demonstrated.
%   \item A cross-individual task of EEG-based emotion recognition is employed to validate the generality of the proposed unified unsupervised framework on the individual difference problem which is common in brain studies.
% \end{itemize}

% \textcolor{black}{The remainder of this paper is structured as following. Section 2 describes the proposed hybrid deep unsupervised EEGFuseNet. Section 3 presents the experimental results of the proposed EEGFuseNet based hypergraph decoding framework on EEG-based emotion recognition. A full comparison with the existing literature is conducted and a full discussion is given. Section 4 summarizes our findings and draws a conclusion of this paper.}

\section{Methodology} \label{sec:Methodology}
\textcolor{black}{In this section, we introduce the proposed EEGFuseNet with the corresponding design and configuration and explain how to efficiently characterize non-stationary high-dimensional EEG signals in an unsupervised manner. An overall EEG preprocessing is first conducted on the collected raw EEG signals to remove noises such as physiological artifacts (e.g. ocular activity and muscle activity) and non-physiological artifacts (e.g. AC electrical and electromagnetic inferences). A full explanation of EEG preprocessing steps is provided in Appendix B of Supplementary Materials. After preprocessing, the EEG data at each trial is further partitioned into a number of segments with a fixed length. A segment-based EEG data is denoted as $X\in{R^{C\times T}}$, representing the signals collected from the electrode channels ($C$) at a period of time points ($T$). Next, $X$ is treated as the input to the proposed hybrid EEGFuseNet and the corresponding deep features are characterized and fused based on unsupervised learning.} The proposed EEGFuseNet mitigates the limitations of the existing state-of-the-art feature extraction and fusion methods and provides a number of practical benefits, for example, easy modification and simple training, for EEG signals collected under different environment variables in various applications. Next, the proposed hybrid deep encoder-decoder network architecture will be illustrated in details. More precisely, we will introduce \textcolor{black}{(1) how to construct the basic architecture of the proposed EEGFuseNet from the classical CNN, (2) how to incorporate GAN into the CNN-based network to generate high-quality features, (3) how to incorporate RNN into the CNN-GAN based network to better fuse both temporal and spatial information and develop the final architecture of EEGFuseNet.}

\subsection{CNN based}
% \subsubsection{CNN based}
\label{sec:CNNbased}
\textcolor{black}{In a typical CNN based deep encoder-decoder network, the encoder consists of convolution layers for extracting useful information from the given input ($X$) and converting it into a single dimensional vector (hidden vector), and the decoder consists of deconvolutional layers for upscaling the encoder feature maps and transferring the hidden vector to the generated output ($Y$) \cite{ye2019understanding}. Through maximizing the similarity between $X$ and $Y$ (in other words, minimizing the loss function given as $\mathcal{L}(X,Y)=\|X-Y\|_2^2$), the autoencoder structure is jointly trained and the learnt hidden vector is considered as an informative latent feature representation of $X$ and used for further data analysis and modelling. \textcolor{black}{Noteworthy, the autoencoder architecture is a self-learning paradigm, which does not require any labeling information during training process and is significantly easier to train comparing to the other common feature extraction architectures \cite{jiao2018deep,chen2019deep}. Thus, it would be suitable to solve the small size problem of EEG data with label missing.}}

\begin{figure}
\begin{center}
\includegraphics[width=0.5\textwidth]{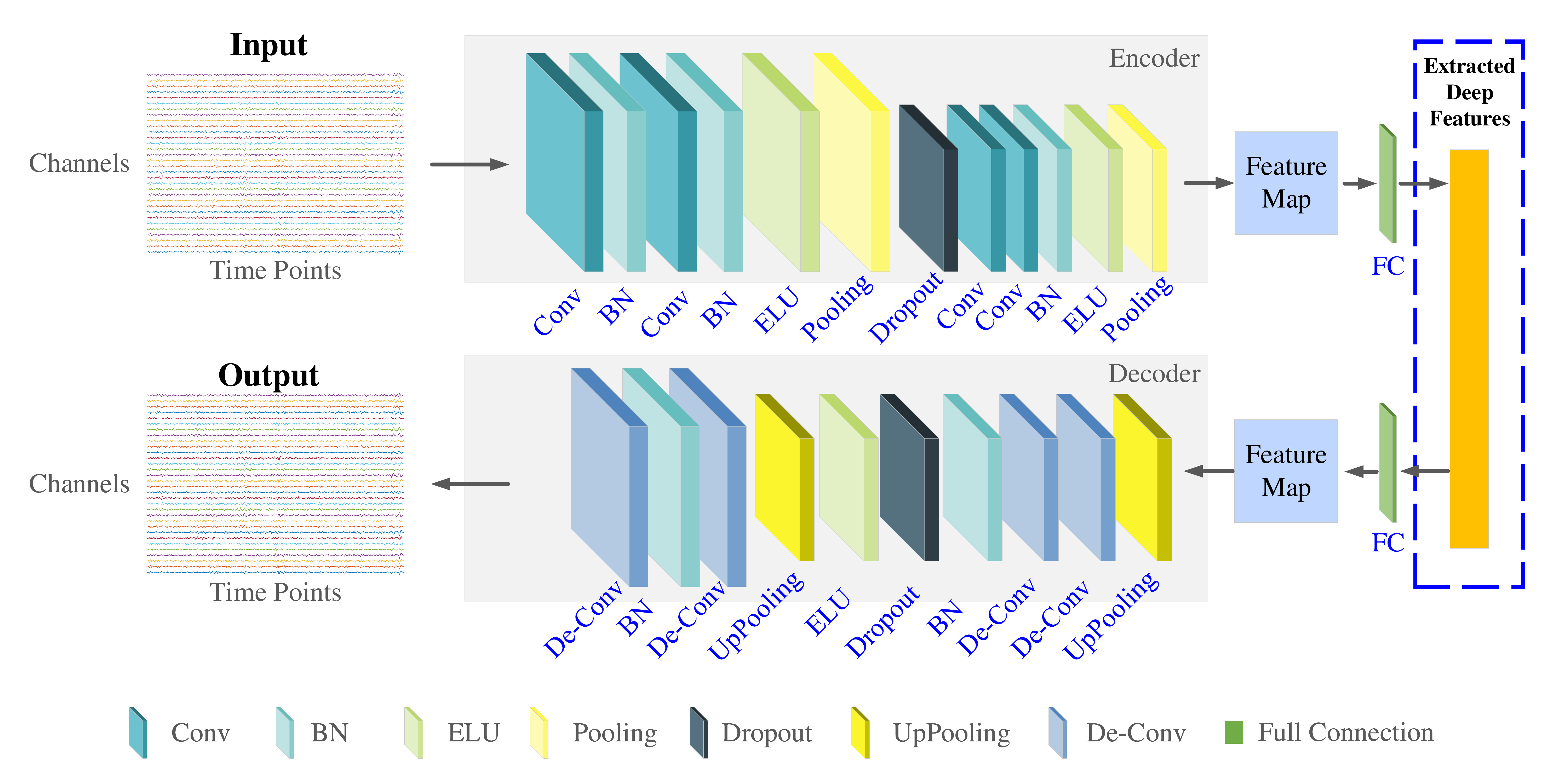}
\end{center}
\caption{An illustration of the CNN based encoder-decoder network.}
\label{fig:CNNflowchart}
\end{figure}

For time-series EEG signals, both spatial and temporal information are important which represents the relationships of the brain activities at different brain locations and the changing dynamics of brain patterns along the time. Inspired from EEGNet structure \cite{lawhern2018eegnet}, a CNN based deep encoder-decoder network is developed as shown in Fig. \ref{fig:CNNflowchart}. A sequential two-dimensional convolutional layers are implemented to generate feature maps covering EEG spatial information at different frequency bands, where the filter length is the half of the sampling rate of input data. Batch normalization (BN) is adopted to normalize each training mini-batch and speed up network training process by reducing internal covariate shift. The activation function, exponential linear units (ELU), is added in convolutional and deconvolutional layers for model fitting improvement. Notably, as the input EEG signals consist of channels and time points ($X\in{R^{C\times T}}$), two-dimensional convolution functions are adopted here, instead of one-dimensional convolution function. In the architecture, the encoder performs convolution and down-sampling, while the decoder performs deconvolution and up-sampling to reconstruct the input EEG signals. \textcolor{black}{The main possible benefits of the multiply convolution layers include: (1) compact, comprehensive and complete EEG pattern characterization from different dimensions; (2) relationship explorations within and between the extracted feature maps and feature fusion in an optimal approach; (3) less parameters to fit with the implementation of subsampling layers. Thus, the design of the convolution layers could be capable of providing an efficient way to learn spatial-temporal dynamics from time-series EEG signals collected at different brain locations and integrate the sample points to a compact and deep feature representation vector which has been demonstrated to be useful for accurate and efficient data description \cite{rifai2011contractive,gehring2013extracting}. This network could offer a baseline for unsupervised deep feature characterization and fusion.} Specifically, the encoder network consists of 4 convolution layers. The weights in the training process are initialized randomly. In the design of an encoder-decoder architecture, each encoder layer would have a corresponding decoder layer. Thus, there are also 4 deconvolution layers in decoder part. The final decoder output is to reconstruct the input EEG signals with minimized difference. In the model training process, we use the mean squared error (MSE) as the objective function to measure the difference between the input EEG signals $X\in{R^{C\times T}}$ and the reconstructed EEG signals $Y\in{R^{C\times T}}$ from the estimated deep features by the network, given as $\textrm{loss}=\|X-Y\|^2_2$. A perfect model would have a loss of 0. The specific architecture details are presented in Appendix C of Supplementary Materials (Table S1). 

% \begin{equation}
% \textrm{loss}=\|X-Y\|^2_2.
% \end{equation}

\subsection{Hybrid CNN-GAN based}
% \subsubsection{Hybrid CNN-GAN based}
\label{sec:CNN-GAN}
A traditional encoder-decoder network is easy to train, but the generated features would be with low quality \cite{akbari2018semi}. Many researches have proven GAN could be capable of generating features with high quality from sequential data \cite{makhzani2015adversarial,chen2018unsupervised,sahu2018adversarial}. \textcolor{black}{A general encoder-decoder pipeline with GAN includes a generator (encoder-decoder network) and a discriminator, where the generator is response to reconstruct EEG signals from the extracted deep features and the discriminator is to distinguish whether the input EEG signals is a fake one generated by the generator or a real one collected from human brain.} To further learn the complex structures of the non-stationary time-series EEG data, a hybrid encoder-decoder architecture incorporating CNN and GAN is developed. On the basis of the CNN-based encoder-decoder network presented in Section \ref{sec:CNNbased}, we further develop a hybrid CNN-GAN based deep encoder-decoder network. In the construction of the hybrid CNN-GAN based network, the generator is CNN-based encoder-decoder network as shown in Fig. \ref{fig:CNNflowchart}. \textcolor{black}{The discriminator architecture is the same as the discriminator used in the final proposed EEGFuseNet (Fig. \ref{fig:CNNRNNGAN}). More specific configurations about the designed discriminator are reported in Appendix C of Supplementary Materials (Table S2).} 

\textcolor{black}{In the training process, the generator $G$ characterizes the latent feature representation ($o$) of the sequential EEG signals $X$ and the discriminator $D(X,G(X))\in[0,1]$ measures the probability that the input (real training sample $X$ or synthesized fake sample $G(X)$ produced by the generator) is real or fake. The objective function in the training process is to build a good $D$ that is capable of discriminating the real sample from the generated fake samples and at the same time develop a good $G$ that can produce a fake sample which is as similar as possible to the real ones (two-player minimax game). \textcolor{black}{In the training process, the discriminator inputs are pairs of $X$ and $G(X)$.} The objective function is given as}

\begin{equation}
\mathcal{L}_{GAN}(G,D)={\E_{X}}{[\log{D(X)}]}+{\E_{X}}{\log{[1-D(G(X))]}},
\end{equation}
where the first part $\log{D(X)}$ is the discriminator output for real sample $X$ and the second part $D(G(X))$ is the discriminator output for the generated fake sample based on the estimated $o$. Together with the objective function of $G$,
\begin{equation}
\mathcal{L}_{1}(G)=\|X-G(X))\|_2^2,
\end{equation}
the overall objective function of the hybrid CNN-GAN network is given as
\begin{equation}
\mathcal{L}=\textrm{arg}\min_{G}\max_{D}{(\mathcal{L}_{GAN}(G,D)+\lambda\mathcal{L}_{1}(G))}.
\end{equation}

\color{black}
\subsection{Hybrid CNN-RNN-GAN based (EEGFuseNet)}
\begin{figure*}
\begin{center}
\includegraphics[width=0.8\textwidth]{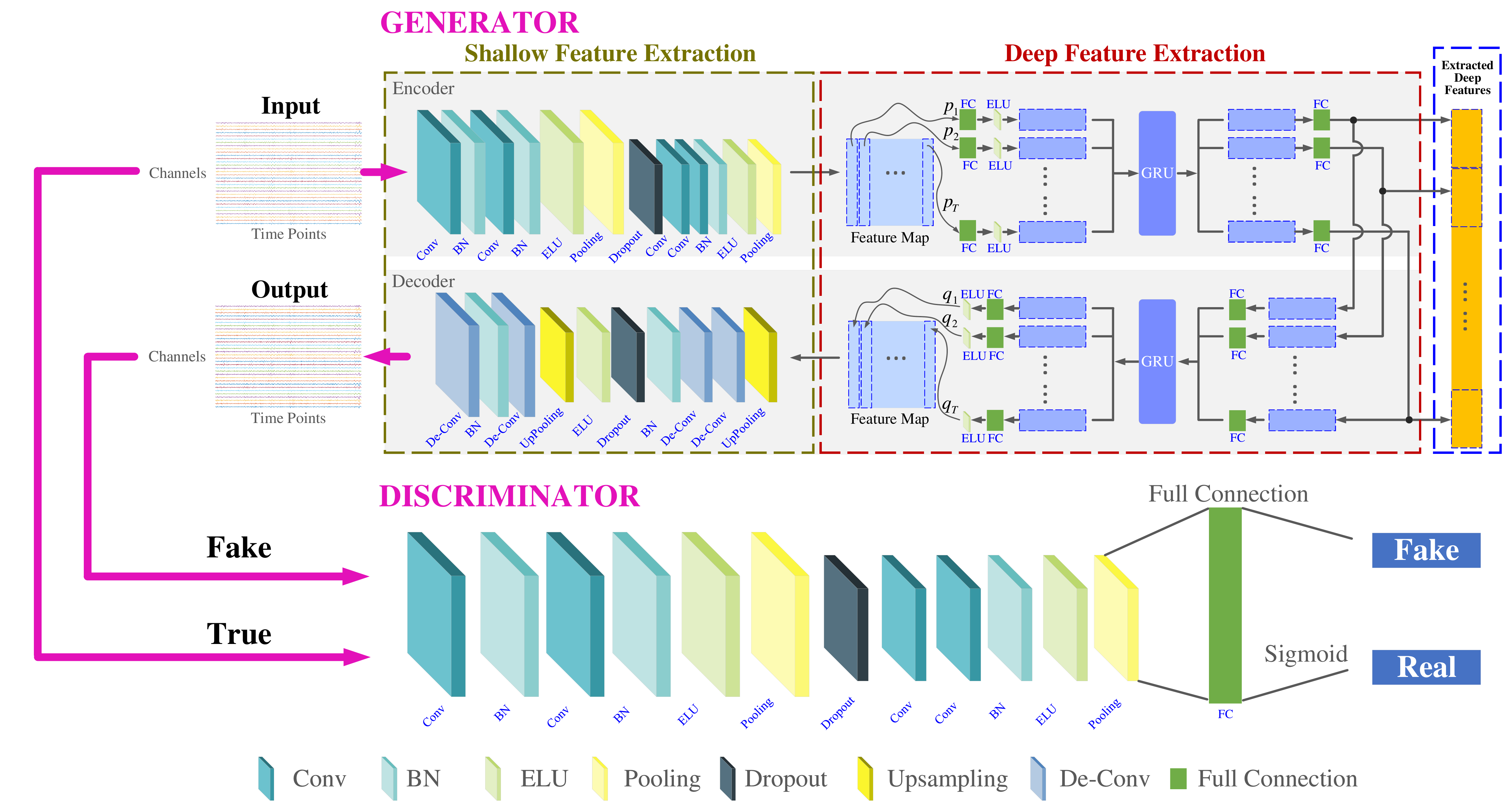}
\end{center}
\caption{\textcolor{black}{The architecture design of the proposed EEGFuseNet. Here, $\{p_1,p_2,...,p_T\}$ and $\{q_1,q_2,...,q_T\}$ are the CNN and reconstructed CNN features extracted from all the channels at each single time point.}}
\label{fig:CNNRNNGAN}
\end{figure*}

% \subsubsection{Hybrid CNN-RNN based}
\label{sec:CNN-RNN-GAN}
\textcolor{black}{According to the nature of EEG signals, there should possess a hierarchical structure with complex dependencies between the extracted features at different time points. The extracted feature at each time point should not be considered as an independent and isolated point.} In the existing works, encoder-decoder models based on RNNs, LSTM and gated recurrent neural networks (GRUs) have recently demonstrated impressive feature characterization performance on sequential data \cite{serban2017hierarchical,vosoughi2016tweet2vec,li2017deep}. To enhance the feature representation of time-series EEG signals, we extend the CNN-GAN based network to a hybrid architecture and extract EEG features by exploiting the advantages of both recurrent and convolutional networks. As shown in Fig. \ref{fig:CNNRNNGAN}, the encoder consists of convolutional layers to extract features from EEG signals at every time point (shallow feature extraction) and the recurrent layers encode the extracted features at every time point to an entire feature representation of the whole input EEG signals (deep feature extraction). The decoder consists of recurrent layers to predict the features at each time point from the output of encoder and deconvolutional layers to reconstruct the features to the original EEG signals. During the EEG processing, the informative features covering both spatial and temporal dynamics are characterized in an effective fusion approach from a joint temporal-spatial domain.

In the hybrid CNN-RNN-GAN based network, the convolution and deconvolution layers in the shallow feature extraction are the same as the designed CNN based network (Fig. \ref{fig:CNNflowchart}). Based on the generated feature maps (the rows and columns refer to the features from channels and time points) in the shallow feature extraction part, the sequential features are characterized in the deep feature extraction part. In the recurrent layers (RNN network), the basic building modules for learning spatial dependencies between neighbors are the LSTM units. Due to the sophisticated training of LSTM, a GRU was proposed \cite{cho2014properties}, which is similar to LSTM that modulates the flow of intimation inside the gating unit without separate memory cell. It has been evident that GRU has shown comparable performance as LSTM on machine learning tasks, with less parameters required \cite{chung2014empirical}. To achieve a higher computation efficiency, we employ a bidirectional GRU in the implementation which is defined as

\begin{equation}
z_t=\sigma(W^{(z)}x_t+U^{(z)}h_{t-1}+b^{(z)},
\end{equation}
\begin{equation}
    r_t=\sigma(W^{(r)}x_t+U^{(r)}h_{t-1}+b^{(r)},
\end{equation}
\begin{equation}
    \hat{h}_t=\phi{(W^{(h)}x_t+U^{(h)}(r_t\odot h_{t-1})+b^{(h)})}
\end{equation}
\begin{equation}
    h_t=(1-z_t)\odot h_{t-1}+z_t\odot \hat{h}_t
\end{equation}
\textcolor{black}{where $x_t$ $(t \in [1,T])$ is the input and $h_t$ $(t \in [1,T])$ is the output. $T$ is the total length of the data. $W^{(z)}$, $W^{(r)}$, $W^{(h)}$, $U^{(z)}$, $U^{(r)}$, and $U^{(h)}$ are weight matrices and $b^{(z)}$, $b^{(r)}$, $b^{(h)}$ are biases, which are learnt in the training process. $z_t$, $r_t$ and $\hat{h}_t$ are update gate vector, reset gate vector and hidden state vector, respectively. $\sigma$ and $\phi$ are sigmoid and tangent function. $\odot$ is an element-wise multiplication. In the implementation, the forward and backward recurrent layers iteratively work on the time point based feature vectors in a sequence and compute the corresponding forward and backward sequences of hidden state vectors. Specifically, the data $\{x_1,x_2,…,x_T\}$ is input to the bidirectional GRU in a forward and backward sequences, respectively. Here, the hidden layer of the forward and backward GRU are denoted as $[h_1^f,h_2^f,…,h_T^f]$ and $[h_1^b,h_2^b,…,h_T^b]$. The outputs of forward and backward GRU at the time point $t$ are given as}
\color{black}
\begin{equation}
    h_t^f=GRU(x_t,h_{t-1}^f),
\end{equation}
\begin{equation}
        h_t^b=GRU(x_t,h_{t-1}^b).
\end{equation}
\textcolor{black}{The output of the bidirection GRU at $t$ point is given as $a_t=h_t^f \oplus h_t^b$, where $\oplus$ indicates vector concatenation. Finally, the generated deep feature representation vector ($o$) is denoted as $o=(a_1,…,a_t,…,a_T)$ and the sequential feature information is captured. For the purpose of EEG feature characterization and fusion, the input EEG signals are first characterized as a sequence of feature vectors at each time point $t$ after the convolutional layers (considered as \textbf{spatial dynamic characterization}) and then sequential features are learnt by recurrent layers to synthesize the past and future dynamic information of time-series EEG signals (considered as \textbf{temporal dynamic characterization}). Thus, the extracted $o$, which is treated as the deep EEG features to be used in the following unsupervised EEG decoding, can represent the entire input EEG signals cross timepoints covering not only the EEG characteristics but also the EEG characteristics in the sequential information.} To improve the implementation efficiency, we update the input of GRU from batch to batch in the training process, where one batch includes a continuous EEG signals at a certain time gap. The specific configurations about the generator in the hybrid CNN-RNN-GAN based encoder-decoder network are presented in Appendix C of Supplementary Materials (Table S3). This architecture successfully fuses the extracted feature representations at different deep levels, at different brain locations, and at different time points, which would be beneficial to the representation of spatial and temporal dynamics in the non-stationary time-series EEG signals.

% \subsection{Hybrid CNN-RNN-GAN based (EEGFuseNet)}
% % \subsubsection{Hybrid CNN-RNN-GAN based}
% \label{sec:CNN-RNN-GAN}
% \textcolor{black}{Based on the network architectures developed above, we finally propose a hybrid deep CNN-RNN-GAN based encoder-decoder network, termed as EEGFuseNet, that incorporates CNN-RNN and GAN networks to realize an efficient and effective deep EEG feature characterization and fusion in an unsupervised manner. Here, the generator is the hybrid CNN-RNN based network presented in section \ref{sec:CNN-RNN}, and the discriminator is the hybrid CNN-GAN based network presented in section \ref{sec:CNN-GAN}.} The architecture design of EEGFuseNet is shown in Fig. \ref{fig:CNNRNNGAN}. The configuration details of the generator and discriminator parts are reported in Appendix C of Supplementary Materials (Table S3 and Table S2). During the EEG processing, the features are characterized in an interactive fusion approach from a joint temporal-spatial domain and the informative and effective features covering both spatial and temporal dynamics are characterized.

\color{black}
\subsection{Unsupervised based Hypergraph Decoding Model}
To solve a pure unsupervised learning based cross-subject EEG-based emotion decoding problem, we introduce the hypergraph theory \cite{zhou2006learning} to realize emotion classification. The EEG samples are treated as vertices and a hypergraph is constructed based on the relationship among these vertices in terms of EEG characteristics. Here, the similarity among the EEG samples is calculated based on the characterized features by EEGFuseNet and the hyperedges are formed to connect a number of EEG samples based on the calculated similarity distributions. Different from a simple graph, a hypergraph is capable of connecting a couple of vertices (more than two) that share similar properties, presenting more general types of relations, and revealing more complex hidden structures than single connections. The emotion classification is realized by partitioning the constructed hypergraph into a specific number of classes, through computing the hypergraph Laplacian and solving it with an optimal eigenspace. The constructed hypergraph is then divided into a number of classes and each class indicates one emotion status. For more details, please refer to Appendix D of Supplementary Materials.

\color{black}
\section{Experimental Results} 
\label{sec:experiment}
The simultaneously recorded EEG signals are used to recognize the corresponding emotion statuses, which has been proved to be effective in tackling with the great deal of complexity and variability in emotions. In this section, we fully evaluate the ability of the proposed EEGFuseNet based hypergraph decoding framework on the application of emotion recognition.

\color{black}
\subsection{Benchmarking}
We conduct extensive experiments on three EEG databases, including DEAP \cite{koelstra2011deap}, MAHNOB-HCI \cite{soleymani2011multimodal}, and SEED\cite{zheng2015investigating}, all of which are commonly used for EEG-based emotion recognition evaluation. In all three databases, the emotions are defined by the dimensional emotion model, i.e. valence, arousal, dominance, liking and predictability. The validity and reliability of the proposed unsupervised framework are fully evaluated, and the performance of unsupervised emotion recognition on different emotion dimensions is carefully quantified and compared with the literature. 

The DEAP database was composed of 32 subjects’ EEG emotion data. For each subject, different specific and strong emotions were evoked by 40 selected music videos, each having a duration of 60s, and the corresponding subjective feedbacks on different emotion dimensions (valence, arousal, dominance, and liking) were given for each music video. EEG signals were recorded at a sampling rate of 512Hz from 32 active AgCl electrode sites according to the international 10-20 system placement. To cross-compare with the other studies, we use a fixed threshold of 5 for each emotion dimension to discretize the subjective feedbacks into two classes (low and high). 
% Therefore, we perform a two-class classification task for model evaluation on DEAP database.

The MAHNOB-HCI database included a total of 30 subjects, whose EEG data were recorded using Biosemi active II system with 32 Ag/AgCl electrodes at a sampling rate of 256Hz. Twenty film clips were selected to evoke emotions and the subjective feedback was given using a score in the range of 1 to 9. In the model evaluation, a fixed threshold of 5 is used to discretize the subjective feedback into binaries for each emotion dimension (valence, arousal, dominance, and predictability). 
% Considering the data completeness, the total sample size are 24 subjects $\times$ 20 videos.

The SEED database included 15 subjects' 62-channel EEG data collected when they were viewing 15 film clips with an average duration around 4 min. The data sampling rate was downsampled to 200Hz. Three emotions were elicited, including negative, neural and positive. 
% The total sample size are 15 subjects $\times$ 15 videos.

\color{black}
\subsection{Experiment Protocols}
To avoid information leaking in the evaluation process, we conduct a leave-one-subject-out cross-validation (LOOCV) subject-independent evaluation protocol on the three databases, where the training and test data are from different subjects and no information overlap exists. Take the DEAP database as an example. We use 31 subjects’ data for training and the remaining 1 subject’s data for testing and repeat the validation process until each subject is treated as the test data for once. In other words, we repeat 32 times and calculate the final cross-validation performance as an average of all the obtained testing results. This validation method provides a fair evaluation of the cross-subject model performance that could accurately estimate the possible recognition accuracies of the newly coming data from new subjects. For comparison with other studies, the decoding performance is evaluated using recognition accuracy $P_{acc}$ and F1-Score $P_f$. Considering the emotion dimensions in the dimensional emotion model are independent to each other, the leave-one-subject-out cross-validation subject-independent evaluation process is separately conducted on each single emotion dimension and the corresponding results are analyzed.

% The total data size are 1280 trials (32 subjects $\times$ 40 videos), and the data dimensionality of each trial is 983040 (32 electrodes $\times$ 512Hz $\times$ 60s). 

\color{black}
\subsection{Network Training}
Not like the media data used in deep learning studies, EEG database is relatively small. The challenge of network training is to well train the network to extract sufficient EEG features and avoid the over-fitting problem. To increase the sample size, each trial is further segmented into a number of segments with a fixed length of 1s. Take DEAP database as an example. Each trial length is 60s, so the number of segments of one trial is equal to 60. Thus, the total sample size is increased from 1280 (32 subjects $\times$ 40 videos) to 76800 (32 subjects $\times$ 40 trials $\times$ 60 segments). In the training process, to avoid information leaking, the segments from one video would all be considered as training data or test data. The weight parameters in convolution layers are initialized with the uniform distribution based on Glorot initialization \cite{glorot2010understanding}. We run 100 training epochs and perform validation stopping. The model weights that generated the lowest validation set loss are saved as the final parameters. The Adam optimizer is with a momentum of 0.9. The mini-batch stochastic gradient descent (SGD) method with a fixed learning rate of 0.001 for generator and of 0.0002 for discriminator is used. Here, the mini-batch size is equal to 128. All the models are trained on an NVIDIA GeForce RTX 2080 GPU, with CUDA 10.0 using the Pytorch API.

\textcolor{black}{On the other hand, due to the computation complexity of the hypergraph construction process, it is very time-consuming to measure the similarity relationships among all the available samples. To improve the computation efficiency in model implementation, we introduce a learning strategy for tackling the computation complexity issue. Specifically, in one round of cross-validation for DEAP database, all the samples from 1 subject are treated as test data (in total 1 subject $\times$ 40 trials $\times$ 60 segments$=$2400 samples) and the samples from the other 31 subjects are used as training data candidates. Then, $\eta\%$ samples are randomly selected from the training data candidates (31 subjects $\times$ 40 trials $\times$ 60 segments$=$74400 samples) and are then used to construct a hypergraph with the test data (2400 samples). Through this approach, the computation efficiency is largely improved. More details about the implementation process are presented in Appendix E of Supplementary Materials. Under a consideration of computation efficiency and performance stability, in the implementation the feature size $\ell$ and hyperedge size $\kappa$ in hypergraph construction and partitioning are set to 64 and 5, while the selected rate $\eta$ in model learning is given to 10.}

\color{black}
\subsection{\textcolor{black}{Evaluation on DEAP Database}}
\textcolor{black}{It is well known that, even evaluating on the same database, different evaluation protocols would lead to a great difference in the results. Generally speaking, the validation methods affect the obtained performance as: \textbf{(1) supervised vs. unsupervised:} supervised methods would have a better result than unsupervised methods, as label information is used for model training in the supervised methods; \textbf{(2) subject-dependent vs. subject-independent:} subject-dependent evaluation methods would have a better performance than subject-independent evaluation methods, as individual difference is not considered in the subject-dependent evaluation methods; \textbf{(3) k-fold CV vs. video-level LOOCV:} k-fold CV methods would have a better performance than video-level LOOCV methods, as it would exist the possibility of having the training and test data from the same video stimulus in the k-fold CV methods; \textbf{(4) video-level LOOCV vs. subject-level LOOCV:} video-level LOOCV methods would have a better performance than subject-level LOOCV methods, as it would exist the possibility of having the training and test data from the same subject in the video-level LOOCV methods.} Table \ref{tab:DEAPCompare} reports the performance comparisons with the existing literature, where the corresponding validation approaches are clearly described. The results show that the proposed EEGFuseNet based hypergraph decoding framework performs a close emotion recognition performance, comparing to the other supervised methods. It is reasonable that unsupervised methods perform poor than supervised methods. For the unsupervised method presented in \cite{liang2019unsupervised}, the original adopted evaluation protocol was subject-dependent leave-one-video-out cross-validation, where cross-subject performance was not considered. To have a fair comparison between our proposed method and Liang \etal's work \cite{liang2019unsupervised}, we evaluate their work using leave-one-subject-out cross-validation subject-independent protocol (same as the one used for our proposed method) and report the results in Table \ref{tab:DEAPCompare}. The comparison results demonstrate our proposed method outperforms Liang \etal’s work, where both $P_{acc}$ and $P_f$ significantly increase from 54.30 to 56.44 and 53.45 to 70.83 for valence, from 55.55 to 58.55 and from 52.77 from 72.00 for arousal, from 57.03 to 61.71 and from 52.84 to 74.32 for dominance, and from 58.91 to 65.89 and from 59.99 to 78.46 for liking. The average increase rates of $P_{acc}$ and $P_{f}$ are 7.35\% and 35.10\%, respectively. The results also reveal that, comparing to handcrafted features \textcolor{black}{that were used in \cite{liang2019unsupervised}}, the characterized deep features by EEGFuseNet could be less sensitive to the individual differences. 

% The validity of recognizing emotional states in four dimensions via subject-independent unsupervised learning method using the characterized deep features by the proposed EEGFuseNet. 

% Pacc: (56.44-54.30)/54.30=0.0394; (58.55-55.55)/55.55=0.0540; (61.71-57.03)/57.03=0.0821;(65.89-58.91)/58.91=0.1185

% Pf: (70.83-53.45)/53.45=0.3252; (72.00-52.77)/52.77=0.3644; (74.32-52.84)/52.84=0.4065; (78.46-59.99)/59.99= 0.3079;

\begin{table}[]
\begin{center}
  \caption{\textcolor{black}{Emotion recognition performance on DEAP database.}}
  \label{tab:DEAPCompare}
  \scalebox{0.7}{
  \color{black}\begin{tabular}{lcccccccc}
    \toprule
     \ \multirow{2}{*}{Methods} & \multicolumn{2}{c}{Valence} & \multicolumn{2}{c}{Arousal} & \multicolumn{2}{c}{Dominance}  & \multicolumn{2}{c}{Liking}\\
     \ & $P_{acc}$ & $P_f$ & $P_{acc}$ & $P_f$ & $P_{acc}$ & $P_f$ & $P_{acc}$ & $P_f$\\
     \midrule
       \ \textbf{\textit{Supervised}} & \multicolumn{4}{l}{\textbf{\textit{Subject-Dependent}}} & \multicolumn{4}{c}{\textbf{\textit{K-fold CV}}}\\
     \midrule
    \  Liu and Sourina\cite{liu2012eeg} &50.80&-&	76.51&- &	-&-&	-&-\\
    \  Li \etal \cite{li2015eeg} & 58.40&-& 	64.20&-& 	65.80&-& 	66.90&-\\
    \ Chen \etal \cite{chen2015electroencephalogram} &67.89&67.83&	69.09&68.96&	-&-&	-&-\\
    \midrule
     \ \textbf{\textit{Supervised}} & \multicolumn{4}{l}{\textbf{\textit{Subject-Dependent}}} & \multicolumn{4}{c}{\textbf{\textit{Video-level LOOCV}}}\\
     \midrule
    \ Koelstra \etal \cite{koelstra2011deap} &57.60&56.30&	62.00&58.30&	-&-&	55.40&50.20 \\
    \ Bahari and Janghorbani \cite{bahari2013eeg} &58.05&-&	64.56&-&	-&-&	67.42&-\\
    \ Naser and Saha \cite{naser2013recognition} & 64.30&-&	66.20&-&	68.90&-&	70.20&- \\
    \ Zhuang \etal \cite{zhuang2017emotion}& 69.10&-&	71.99&-&	-&-&	-&-\\
    \midrule
     \ \textbf{\textit{Supervised}} & \multicolumn{4}{l}{\textbf{\textit{Subject-Independent}}} & \multicolumn{4}{c}{\textbf{\textit{K-fold CV}}}\\
     \midrule
    \ Torres-Valencia \etal \cite{torres2014comparative}& 58.75&-&	55.00&-&	-&-&	-&-\\
    \ Atkinson and Campos \cite{atkinson2016improving}& 73.14&-&	73.06&-&	-&-&	-&-\\
    \ Liu \etal \cite{liu2016emotion} &69.90&-&	71.20&-&	-&-&	-&-\\
    \midrule
      \ \textbf{\textit{Supervised}} & \multicolumn{4}{l}{\textbf{\textit{Subject-Independent}}} & \multicolumn{4}{c}{\textbf{\textit{Subject-level LOOCV}}}\\
      \midrule
    \ Shahnaz \etal \cite{shahnaz2016emotion} &64.71	&74.94&66.51&76.68&	66.88&76.67&	70.52&81.94\\
    \ Song \etal \cite{song2018eeg} &59.29	&-&61.10&-&	-	&-&-&-\\
    \ Chen \etal  \cite{chen2019hierarchical} &67.90&-&	66.50&-&	-&-&	-&-\\
    \ Zhong \etal \cite{zhong2020cross}& 66.23	&-&68.50&-&	-&-&	-&-\\
    \ Du \etal \cite{du2020efficient} &69.06&-&	72.97&-&	-&-&	-&-\\
    \midrule
    \ \textbf{\textit{Unsupervised}} & \multicolumn{4}{l}{\textbf{\textit{Subject-Dependent}}} & \multicolumn{4}{c}{\textbf{\textit{Video-level LOOCV}}}\\
    \midrule
    \ Liang \etal \cite{liang2019unsupervised}  &56.25 &61.25&	62.34 &60.44&	64.22 &64.80&	66.09&77.52\\
    \midrule
    \ \textbf{\textit{Unsupervised}} & \multicolumn{4}{l}{\textbf{\textit{Subject-Independent}}} & \multicolumn{4}{c}{\textbf{\textit{Subject-level LOOCV}}}\\
    \midrule
    \ Liang \etal \cite{liang2019unsupervised}  &54.30 &  53.45&	55.55 & 52.77 &	57.03 & 52.84  &58.91& 59.99\\
    \ \textbf{Proposed Method} & \textbf{56.44}	&\textbf{70.83}&\textbf{58.55}&\textbf{72.00}&	\textbf{61.71}&\textbf{74.32}&	\textbf{65.89}&\textbf{78.46}\\
    \bottomrule
  \end{tabular}
  }
  \end{center}
\end{table}

\subsection{\textcolor{black}{Evaluation on MAHNOB-HCI Database}}
We perform a binary classification task to evaluate the emotion recognition performance of each emotion dimension on MAHNOB-HCI database. The emotion recognition performance using leave-one-subject-out cross-validation subject-independent protocol is summarized in Table \ref{tab:HCIcompare}. The results show our proposed method achieve a comparable performance comparing to the other supervised methods, where the recognition accuracies ($P_{acc}$) of valence, arousal, dominance, and predictability are 60.64$\%$, 62.06$\%$, 67.08$\%$ and 74.63$\%$ and the corresponding F1-Scores ($P_{f}$) are 72.18$\%$, 62.05$\%$, 76.65$\%$, and 83.61$\%$.

\begin{table}[]
\begin{center}
  \caption{\textcolor{black}{Emotion recognition performance on MAHNOB-HCI database.}} 
  \label{tab:HCIcompare}
  \scalebox{0.77}{
  \color{black}\begin{tabular}{lcccccccc}
    \toprule
     \  \multirow{2}{*}{Methods} &\multicolumn{2}{c}{Valence} & \multicolumn{2}{c}{Arousal} & \multicolumn{2}{c}{Dominance}  & \multicolumn{2}{c}{Predictability} \\
     \ &  $P_{acc}$ & $P_f$ & $P_{acc}$ & $P_f$ & $P_{acc}$ & $P_f$ & $P_{acc}$ & $P_f$ \\
     \midrule
        \ \textbf{\textit{Supervised}} & \multicolumn{4}{l}{\textbf{\textit{Subject-Dependent}}} & \multicolumn{4}{c}{\textbf{\textit{Video-level LOOCV}}}\\
     \midrule
         \ Zhu \etal \cite{zhu2014emotion} & 55.72 & 51.44&60.23 &57.77  & - &-&-&- \\
         \midrule
        \ \textbf{\textit{Supervised}} & \multicolumn{4}{l}{\textbf{\textit{Subject-Independent}}} & \multicolumn{4}{c}{\textbf{\textit{Subject-level LOOCV}}}\\
        \midrule
    \ Soleymani \etal \cite{soleymani2011multimodal} & 57.00 & 56.00 &52.40 &42.00& - &-&-&- \\
    \ Huang \etal \cite{huang2016multi} & 62.13 &  - & 61.80 &-& - &-&-&-\\
    \ Yin \etal \cite{yin2020locally} &  69.93 & 71.22 & 67.43 &68.58 & - &-&-&- \\ 
    \midrule
     \ \textbf{\textit{Unsupervised}} & \multicolumn{4}{l}{\textbf{\textit{Subject-Independent}}} & \multicolumn{4}{c}{\textbf{\textit{Subject-level LOOCV}}}\\
    \midrule
    \ \textbf{Proposed Method} & 60.64 & 72.18  & 62.06	 & 62.05  & 67.08	 & 76.65  & 74.63	& 83.61 \\
    \bottomrule
  \end{tabular}
  }
  \end{center}
\end{table}

% \begin{table*}[]
% \begin{center}
%   \caption{\textcolor{black}{Emotion recognition performance on MAHNOB-HCI database.}} 
%   \label{tab:HCIcompare}
%   \scalebox{1}{
%   \color{black}\begin{tabular}{lcccccccc}
%     \toprule
%      \  \multirow{2}{*}{Methods} &\multicolumn{2}{c}{Valence} & \multicolumn{2}{c}{Arousal} & \multicolumn{2}{c}{Dominance}  & \multicolumn{2}{c}{Predictability} \\
%      \ &  $P_{acc}$ & $P_f$ & $P_{acc}$ & $P_f$ & $P_{acc}$ & $P_f$ & $P_{acc}$ & $P_f$ \\
%      \midrule
%         \ \textbf{\textit{Supervised}} & \multicolumn{4}{l}{\textbf{\textit{Subject-Dependent}}} & \multicolumn{4}{c}{\textbf{\textit{Video-level LOOCV}}}\\
%      \midrule
%          \ Zhu \etal \cite{zhu2014emotion} & 55.72 & 51.44&60.23 &57.77  & - &-&-&- \\
%          \midrule
%         \ \textbf{\textit{Supervised}} & \multicolumn{4}{l}{\textbf{\textit{Subject-Independent}}} & \multicolumn{4}{c}{\textbf{\textit{Subject-level LOOCV}}}\\
%         \midrule
%     \ Soleymani \etal \cite{soleymani2011multimodal} & 57.00 & 56.00 &52.40 &42.00& - &-&-&- \\
%     \ Huang \etal \cite{huang2016multi} & 62.13 &  - & 61.80 &-& - &-&-&-\\
%     \ Yin \etal \cite{yin2020locally} &  69.93 & 71.22 & 67.43 &68.58 & - &-&-&- \\ 
%     \midrule
%      \ \textbf{\textit{Unsupervised}} & \multicolumn{4}{l}{\textbf{\textit{Subject-Independent}}} & \multicolumn{4}{c}{\textbf{\textit{Subject-level LOOCV}}}\\
%     \midrule
%     \ Proposed Method& 60.64 & 72.18  & 62.06	 & 62.05  & 67.08	 & 76.65  & 74.63	& 83.61 \\
%     \bottomrule
%   \end{tabular}
%   }
%   \end{center}
% \end{table*}

\subsection{\textcolor{black}{Evaluation on SEED Database}}
For the model evaluation on SEED database, two classification tasks are performed: 2-class (an emotion recognition of negative and positive) and 3-class (an emotion recognition of negative, neural and positive). A subject-independent leave-one-subject-out cross-validation evaluation method is used and the corresponding results are summarized in Table \ref{tab:seedCompare}. In the model comparison, the results show the supervised model with transfer learning strategy (both training and test data are used for model learning) perform the best, where the 3-class classification accuracy is around 80$\%$. Our proposed unsupervised method achieves a comparable result comparing to the supervised model without transfer learning strategy (only training data is used for model learning), where the 3-class classification accuracy is 59.06$\%$ for unsupervised method and 58.23$\%$ for supervised method. \textcolor{black}{As most of current studies on SEED database are based on transfer learning strategy, the presented supervised result without transfer learning strategy was a baseline method reported in \cite{li2019domain} for comparing the performance with and without transfer learning strategy.}

\begin{table}[]
\begin{center}
  \caption{\textcolor{black}{Emotion recognition performance on SEED database.}}
  \label{tab:seedCompare}
  \scalebox{0.77}{
  \color{black}\begin{tabular}{lccc}
    \toprule
      \ Methods & Classification Task & $P_{acc}$ & $P_{f}$ \\
       \midrule
           \ \textbf{\textit{Supervised}} & \multicolumn{2}{l}{\textbf{\textit{Subject-Dependent}}} & \multicolumn{1}{c}{\textbf{\textit{Video-level LOOCV}}}\\
     \midrule
     \ Zheng and Lu \cite{zheng2015investigating} & 3-Class & 86.08 & - \\
     \ Zheng \cite{zheng2016multichannel} & 3-Class & 82.96 & - \\
     \ Li \etal \cite{li2018novel} & 3-Class & 92.38 & - \\
      \ Song \etal \cite{song2018eeg} & 3-Class & 90.40 & - \\
     \ Li \etal \cite{li2019regional} & 3-Class & 93.38 & - \\
        \midrule
       \ \textbf{\textit{Supervised with Transfer Learning}} & \multicolumn{2}{l}{\textbf{\textit{Subject-Independent}}} & \multicolumn{1}{c}{\textbf{\textit{Subject-level LOOCV}}}\\
        \midrule
        \ Pan \etal \cite{pan2010domain} & 3-Class & 63.64 & - \\
         \ Li \etal \cite{li2018novel} & 3-Class & 83.28 & - \\
         \ Song \etal \cite{song2018eeg} & 3-Class & 79.95 & - \\
        \ Li \etal \cite{li2019regional} & 3-Class & 84.16 & - \\
        \ Li \etal \cite{li2019domain} & 3-Class & 88.28 & - \\
        \midrule
         \ \textbf{\textit{Supervised without Transfer Learning}} & \multicolumn{2}{l}{\textbf{\textit{Subject-Independent}}} & \multicolumn{1}{c}{\textbf{\textit{Subject-level LOOCV}}}\\
          \midrule
         \ Li \etal \cite{li2019domain} (source domain only) & 3-Class & 58.23 & - \\
        \midrule
        \ \textbf{\textit{Unsupervised}} & \multicolumn{2}{l}{\textbf{\textit{Subject-Independent}}} & \multicolumn{1}{c}{\textbf{\textit{Subject-level LOOCV}}}\\
        \midrule
     \ \multirow{2}{*}{\textbf{Proposed Method}} & {2-Class} & 80.83 & 82.03  \\
     \ & {3-Class} & 59.06 & -\\
    \bottomrule
  \end{tabular}
  }
  \end{center}
\end{table}

\color{black}
\section{Discussion and Conclusion} 
To fully study the model performance, we compare our proposed method with the existing feature representation methods and decoding methods. Also, we conduct an ablation study to show the effectiveness of the modules in the feature characterization and fusion scheme in EEGFuseNet. Besides $P_{acc}$ and $P_f$, we introduce the normalized mutual information (NMI) as another performance metric to compare the evaluation performance and check the corresponding clustering quality. It is worth note that all the model performance evaluations are conducted using leave-one-subject-out cross-validation subject-independent evaluation protocol.

\subsection{Performance Comparison with Different Feature Representation Methods}
\label{FeaturePerformance}
We compare the feature characterization ability of EEGFuseNet with the commonly used traditional EEG features in the literature, such as time domain features, power spectral features, and differential entropy features. Time domain features characterize the statistical patterns, Hjorth features and shape information of time-series EEG data; power spectral features characterize the spectral powers at different frequency bands; and differential entropy features characterize the differential entropy at different frequency bands. More details about the traditional EEG feature characterization can be found in Appendix F of Supplementary Materials. To make the results comparable, the extracted traditional features and the EEGFuseNet features are mapped to the same feature dimensionality and input to the hypergraph decoding model to realize emotion recognition. The parameters in the hypergraph decoding model are the same as the proposed method, where $\kappa$ and $\eta$ values are set to 5 and 10, respectively. We evaluate the performance on three public databases and present the results in Table \ref{tab:traditionFeaDEAP}, Table \ref{tab:traditionFeaHCI} and Table \ref{tab:traditionFeaSEED}. The results indicate that the EEGFuseNet features achieve the best performance on all three databases compared with time domain features, power spectral features, and differential entropy features. Both $P_{acc}$ and $P_f$ values of EEGFuseNet are significantly higher than the traditional features on all three databases. The NMI values of EEGFuseNet are higher than the traditional features on MAHNOB-HCI and SEED databases. However, for DEAP database, the obtained NMI values of EEGFuseNet are lower than that of traditional features. One possible reason could be that the biased sample distribution problem of low and high classes in DEAP database makes the unstable and unexpectable results in NMI values. The above results demonstrate EEGFuseNet achieves an overall better performance that the other feature representation methods, integrating both spatial and temporal dynamic characteristics in EEG signals.

% One possible reason could be that DEAP database has a biased sample distribution problem of low and high classes in the emotion dimensions and the calculation of NMI value is very sensitive to the unbalanced prediction issues.

\begin{table*}[]
\begin{center}
  \caption{\textcolor{black}{Emotion recognition performance with traditional EEG features using leave-one-subject-out cross-validation subject-independent protocol on DEAP database.}}
  \label{tab:traditionFeaDEAP}
  \scalebox{1}{
  \color{black}\begin{tabular}{lcccccccccccc}
    \toprule
                \ \multirow{2}{*}{Methods} & \multicolumn{3}{c}{Valence} & \multicolumn{3}{c}{Arousal} & \multicolumn{3}{c}{Dominance}   & \multicolumn{3}{c}{Liking} \\
      \  & $P_{acc}$ & $P_f$ & NMI & $P_{acc}$ & $P_f$ & NMI & $P_{acc}$ & $P_f$ & NMI & $P_{acc}$ & $P_f$  & NMI\\
    \midrule
    \ {Time domain features}& 54.47  &68.24 &0.0144	&54.93 & 68.55& 0.0083& 59.15&71.25 &0.0107&62.38	 &75.24& 0.0125\\
    \  {Power spectral features}  &54.52 &68.50 &0.0198&56.03& 69.34&0.0244 &57.48& 71.36&0.0234 &60.66&  73.31&0.0227\\
    \ {Differential entropy features} &54.14& 68.81	&0.0189& 56.54 &70.05 &0.0164	&57.45&71.60& 0.0177	&63.35&76.16 &0.0125\\
    \ {\textbf{Proposed Method}} & \textbf{56.44} & \textbf{70.83}	&0.0013& \textbf{58.55}  & \textbf{72.00}&	0.0011&\textbf{61.71}&\textbf{74.32}& 0.0014	&\textbf{65.89}&\textbf{78.46} &0.0010 \\
    \bottomrule
  \end{tabular}
  }
  \end{center}
\end{table*}

\begin{table*}[]
\begin{center}
  \caption{\textcolor{black}{Emotion recognition performance with traditional EEG features using leave-one-subject-out cross-validation subject-independent protocol on MAHNOB-HCI database.}}
  \label{tab:traditionFeaHCI}
  \scalebox{1}{
  \color{black}\begin{tabular}{lcccccccccccc}
    \toprule
                \ \multirow{2}{*}{Methods} & \multicolumn{3}{c}{Valence} & \multicolumn{3}{c}{Arousal} & \multicolumn{3}{c}{Dominance}   & \multicolumn{3}{c}{Predictability} \\
      \  & $P_{acc}$ & $P_f$ & NMI & $P_{acc}$ & $P_f$ & NMI & $P_{acc}$ & $P_f$ & NMI & $P_{acc}$ & $P_f$  & NMI\\
    \midrule
    \ {Time domain features}&53.70&69.36&0&66.18 &54.70&0.0316&59.83&73.41&0&69.45&81.07& 0.0001\\
    \  {Power spectral features}&53.76&69.37&0.0001&65.42&53.10&0.0181&59.74&73.46&0.0001&69.36&81.05 &0\\
    \ {Differential entropy features}&54.12&69.47&0.0050&65.11&56.65&0.0249&60.11&73.62&0.0068&69.78&81.11 &0.0031 \\
    \ {\textbf{Proposed Method}} &\textbf{60.64}&\textbf{72.18}&\textbf{0.1129}&62.06&\textbf{62.05}&\textbf{0.0918}&\textbf{67.08}&\textbf{76.65}&\textbf{0.1783}&\textbf{74.63}&\textbf{83.61}&\textbf{0.1829}\\
    \bottomrule
  \end{tabular}
  }
  \end{center}
\end{table*}

\begin{table}[]
\begin{center}
  \caption{\textcolor{black}{Emotion recognition performance with traditional EEG features using leave-one-subject-out cross-validation subject-independent protocol on SEED database.}}
  \label{tab:traditionFeaSEED}
  \scalebox{0.9}{
  \color{black}\begin{tabular}{lccccc}
    \toprule
    \  \multirow{2}{*}{Methods} & \multicolumn{3}{c}{Two-Class} & \multicolumn{2}{c}{Three-Class} \\
    \ & $P_{acc}$ & $P_f$ &  NMI & $P_{acc}$ & NMI \\
    \midrule
     \ Time domain features &  65.13 & 71.09 & 0.1748 & 47.03&0.1742\\
    \ Power spectral features & 70.08&75.86 &0.2869 & 49.65 & 0.2745\\
     \ Differential entropy features  & 64.39&67.82 &0.1745 &49.88 &0.1994\\
      \ \textbf{Proposed Method} & \textbf{80.83} & \textbf{82.03} & \textbf{0.4381} & \textbf{59.06} & \textbf{0.3569}\\
    \bottomrule
  \end{tabular}
  }
  \end{center}
\end{table}

% \begin{table*}[]
% \begin{center}
%   \caption{\textcolor{black}{Emotion recognition performance with traditional EEG features using leave-one-subject-out cross-validation subject-independent protocol on SEED database.}}
%   \label{tab:traditionFeaSEED}
%   \scalebox{1}{
%   \color{black}\begin{tabular}{lccccc}
%     \toprule
%     \  \multirow{2}{*}{Methods} & \multicolumn{3}{c}{Two-Class} & \multicolumn{2}{c}{Three-Class} \\
%     \ & $P_{acc}$ & $P_f$ &  NMI & $P_{acc}$ & NMI \\
%     \midrule
%      \ Time domain features &  65.13 & 71.09 & 0.1748 & 47.03&0.1742\\
%     \ Power spectral features & 70.08&75.86 &0.2869 & 49.65 & 0.2745\\
%      \ Differential entropy features  & 64.39&67.82 &0.1745 &49.88 &0.1994\\
%       \ Proposed Method & \textbf{80.83} & \textbf{82.03} & \textbf{0.4381} & \textbf{59.06} & \textbf{0.3569}\\
%     \bottomrule
%   \end{tabular}
%   }
%   \end{center}
% \end{table*}

\color{black}
\subsection{Performance Comparison with Different Decoding Models}
\label{modelPerformance}
We evaluate the robustness of the proposed unsupervised hypergraph decoding model by comparing to the state-of-the-art decoding methods. For example, simple graph based method, principal component analysis (PCA) and K-means clustering method (PCA+Kmeans), K-nearest neighbors (KNN) algorithm, robust continuous clustering method (RCC) \cite{shah2017robust}, and directed graph based agglomerative algorithm (AGDL) \cite{zhang2012graph}. PCA+Kmeans and KNN are two baseline methods. Simple graph works with pair-wise relationship measurement. RCC is a clustering algorithm by optimizing a continuous objective based on robust estimation. AGDL is an agglomerative clustering method based on a direct graph, where the product of average indegree and average outdegree was measured to guarantee the stability of cluster results. Another recently popular unsupervised method presented by Yang \etal \cite{yang2016joint} was not included in this comparison, because Yang \etal's work was an image-based end-to-end learning framework to learn effective features and implement clusters jointly, which is hard to separate the clustering part from the framework and directly extend to EEG tasks. To make the results comparable, the used EEG features are extracted from the proposed EEGFuseNet with same parameter settings. The performance comparisons are conducted on all three databases, and the corresponding results are reported in Table \ref{tab:clusterModelDEAP}, Table \ref{tab:clusterModelHCI}, and Table \ref{tab:clusterModelSEED}. Through comparing the emotion recognition performance on different emotion dimensions and different databases, the results show that our proposed unsupervised hypergraph decoding model achieves the most robust results across different subjects, different trials and different experimental environments. Comparing to pair-wise relationship measurement in simple graph, hypergraph construction and partitioning could be more beneficial to describe the complex hidden relationships of EEG data in decoding problems. For the other decoding methods, the recognition results are similar to simple graph's. These results verify a simple unsupervised method is not suitable for solving the complex and difficult decoding problems using high-dimensional EEG signals.

 \begin{table*}[]
\begin{center}
  \caption{\textcolor{black}{Emotion recognition performance with the state-of-the-art decoding methods using leave-one-subject-out cross-validation subject-independent protocol on DEAP database.}}
  \label{tab:clusterModelDEAP}
  \scalebox{0.9}{
  \color{black}\begin{tabular}{lcccccccccccc}
    \toprule
                \ \multirow{2}{*}{Methods} & \multicolumn{3}{c}{Valence} & \multicolumn{3}{c}{Arousal} & \multicolumn{3}{c}{Dominance}   & \multicolumn{3}{c}{Liking} \\
      \  & $P_{acc}$ & $P_f$ & NMI & $P_{acc}$ & $P_f$ & NMI & $P_{acc}$ & $P_f$ & NMI & $P_{acc}$ & $P_f$  & NMI\\
    \midrule
     \ Simple graph &55.05 &68.55 &0.0001 & 57.38 &69.62&0.0001 &60.21&72.00&0.0001&63.39&75.67&0.0001\\
    \ PCA+k-means &51.66&59.01&0.0034 &52.52&59.80&0.0018&53.15&61.03&0.0026&55.14&64.81&0.0032\\
    \ KNN & 50.30&54.64&0.0003&51.56&59.05&0.0004&53.87&61.00&0.0005&59.51&71.82&0.0004\\
    \ RCC \cite{shah2017robust} & 55.31 &69.52&0.0007 & 57.45	&70.74&0.0005 & 60.35	&73.22&0.0006& 64.71 &77.50&0.0005\\
    \ AGDL \cite{zhang2012graph} & 51.43 & 55.58&0.0004& 50.83&55.45&0.0004	 &  50.12&56.52 &0.0004	&  52.71 &60.25  &0.0005\\
    \ \textbf{Proposed Method}& \textbf{56.44} & \textbf{70.83}	&0.0013& \textbf{58.55}  & \textbf{72.00}&	0.0011&\textbf{61.71}&\textbf{74.32}& 0.0014	&\textbf{65.89}&\textbf{78.46} &0.0010 \\
    \bottomrule
  \end{tabular}
  }
  \end{center}
\end{table*}

 \begin{table*}[]
\begin{center}
  \caption{\textcolor{black}{Emotion recognition performance with the state-of-the-art decoding methods using leave-one-subject-out cross-validation subject-independent protocol on MAHNOB-HCI database.}}
  \label{tab:clusterModelHCI}
  \scalebox{0.9}{
  \color{black}\begin{tabular}{lcccccccccccc}
    \toprule
                \ \multirow{2}{*}{Methods} & \multicolumn{3}{c}{Valence} & \multicolumn{3}{c}{Arousal} & \multicolumn{3}{c}{Dominance}   & \multicolumn{3}{c}{Predictability} \\
      \  & $P_{acc}$ & $P_f$ & NMI & $P_{acc}$ & $P_f$ & NMI & $P_{acc}$ & $P_f$ & NMI & $P_{acc}$ & $P_f$  & NMI\\
    \midrule
     \ Simple graph &55.05	&67.34&0.0031&50.84&64.69&0.0029&59.80&73.47&0.0022&	69.36&81.05&0\\
    \ PCA+k-means &53.70&69.36&0	&55.86&46.09&0.0209& 59.68&73.44&0&69.36&81.05&0\\
    \ KNN & 55.17 &56.39&0.0073&54.94 &54.39&0.0057&59.13&66.73&0.0065&67.79&79.03&0.0073\\
    \ RCC \cite{shah2017robust} &62.27&70.02&0.0657 & 59.89 &64.13&0.0484& 67.71 &75.19&0.0761&72.74&81.62&0.0841\\
    \ AGDL \cite{zhang2012graph} & 53.72 &69.37&0.0005& 50.62 &64.65&0.0005& 59.69 &73.44&0.0004& 69.38 &81.06&0.0005\\
    \ \textbf{Proposed Method} & 60.64 & \textbf{72.18} & \textbf{0.1129} & \textbf{62.06} & 62.05 &\textbf{0.0918} &67.08 &\textbf{76.65} &\textbf{0.1783} &\textbf{74.63} &\textbf{83.61} &\textbf{0.1829}\\
    \bottomrule
  \end{tabular}
  }
  \end{center}
\end{table*}

\begin{table}[]
\begin{center}
  \caption{\textcolor{black}{Emotion recognition performance with the state-of-the-art decoding methods using leave-one-subject-out cross-validation subject-independent protocol on SEED database.}}
  \label{tab:clusterModelSEED}
  \scalebox{0.9}{
  \color{black}\begin{tabular}{lccccc}
    \toprule
    \  \multirow{2}{*}{Methods} & \multicolumn{3}{c}{Two-Class} & \multicolumn{2}{c}{Three-Class} \\
    \ & $P_{acc}$ & $P_f$ &  NMI & $P_{acc}$ & NMI \\
    \midrule
    \ Simple graph & 64.16 &64.15 &0.1728& 35.82 &0.0018\\
    \ PCA+k-means & 56.12 & 65.27&0.0658& 40.09 &0.0686\\
    \ KNN & 60.87 &39.18 &0.1139& 42.04 &0.1132\\
    \ RCC \cite{shah2017robust} & 55.52&  66.24&0.0601&  38.27 &0.0752\\
    \ AGDL \cite{zhang2012graph} & 51.09  &67.63&0& 34.47 &0\\
    \ \textbf{Proposed Method} &\textbf{80.83} & \textbf{82.03} & \textbf{0.4381} & \textbf{59.06}&\textbf{0.3569}\\
    \bottomrule
  \end{tabular}
  }
  \end{center}
\end{table}

% \begin{table*}[]
% \begin{center}
%   \caption{\textcolor{black}{Emotion recognition performance with classic unsupervised methods using leave-one-subject-out cross-validation subject-independent protocol on SEED database.}}
%   \label{tab:clusterModelSEED}
%   \scalebox{1}{
%   \color{black}\begin{tabular}{lccccc}
%     \toprule
%     \  \multirow{2}{*}{Methods} & \multicolumn{3}{c}{Two-Class} & \multicolumn{2}{c}{Three-Class} \\
%     \ & $P_{acc}$ & $P_f$ &  NMI & $P_{acc}$ & NMI \\
%     \midrule
%     \ Simple graph & 64.16 &64.15 &0.1728& 35.82 &0.0018\\
%     \ PCA+k-means & 56.12 & 65.27&0.0658& 40.09 &0.0686\\
%     \ KNN & 60.87 &39.18 &0.1139& 42.04 &0.1132\\
%     \ RCC \cite{shah2017robust} & 55.52&  66.24&0.0601&  38.27 &0.0752\\
%     \ AGDL \cite{zhang2012graph} & 51.09  &67.63&0& 34.47 &0\\
%     \ Proposed Method &\textbf{80.83} & \textbf{82.03} & \textbf{0.4381} & \textbf{59.06}&\textbf{0.3569}\\
%     \bottomrule
%   \end{tabular}
%   }
%   \end{center}
% \end{table*}

\color{black}
\subsection{Ablation Study}
The effectiveness of each component in our proposed EEGFuseNet is fully validated in an ablation study based on three different databases. EEGFuseNet is built upon CNN-based encoder-decoder with two additional modules: GAN for model performance enhancement and RNN for temporal feature dynamic measurement. In the ablation study, we compare our proposed EEGFuseNet with three variant models:

\begin{itemize}
  \item CNN based: only the basic CNN based encoder-decoder;
  \item CNN-GAN based: the CNN based encoder-decoder with only GAN module;
  \item CNN-RNN based: the CNN based encoder-decoder with only RNN module.
\end{itemize}

The corresponding unsupervised based emotion recognition performance of ablation study on different emotion dimensions and three different databases are reported in Table \ref{tab:diffNetworkDEAP}, Table \ref{tab:diffNetworkHCI}, and Table \ref{tab:diffNetworkSEED}. Note here the performance is only affected by the characterized EEG features using different network configurations, where the utilized hypergraph decoding model is the same. The results show that EEGFuseNet outperforms these variant models, which is more capable of characterizing and fusing emotion related deep EEG features in a high quality and achieving better cross-subject based emotion recognition performance. Besides, the benefit of hybrid methods is also demonstrated. The hybrid networks (CNN-GAN based, CNN-RNN based, and CNN-RNN-GAN based) outperform the single model (CNN based). These results show hybrid networks are more flexible and stable to handle the data diversity issue and are more beneficial to high-quality EEG feature characterization and information fusion across spatial and temporal dynamics.

\begin{table*}[]
\begin{center}
  \caption{\textcolor{black}{Emotion recognition performance comparison with different network configurations using leave-one-subject-out cross-validation subject-independent protocol on DEAP database.}}
  \label{tab:diffNetworkDEAP}
  \scalebox{0.9}{
  \color{black}\begin{tabular}{lcccccccccccc}
    \toprule
           \ \multirow{2}{*}{Networks} & \multicolumn{3}{c}{Valence} & \multicolumn{3}{c}{Arousal} & \multicolumn{3}{c}{Dominance}   & \multicolumn{3}{c}{Liking} \\
      \  & $P_{acc}$ & $P_f$ &NMI & $P_{acc}$ & $P_f$ &NMI & $P_{acc}$ & $P_f$ &NMI & $P_{acc}$ & $P_f$ &NMI \\
    \midrule
    \ CNN based &54.78  &65.48 	&0.0012&56.63	 & 66.51 &0.0013 &57.31 &68.17&0.0010&62.37&72.33 &0.0012\\
    \ CNN-GAN based &55.18 &68.95    & 0.0008  &57.71 &70.08 &0.0010 &	60.27 &72.43& 0.0018&63.56&76.12&0.0009\\
    \ CNN-RNN based & 55.34          &69.06    &  0.0013  &56.84 &69.98 &0.0013&	59.78  &72.27& 0.0008 &64.49&76.49&0.0010\\
     \ \textbf{CNN-RNN-GAN based (EEGFuseNet)} & \textbf{56.44}	&\textbf{70.83}& \textbf{0.0013} &\textbf{58.55}&\textbf{72.00}& 0.0011	&\textbf{61.71}&\textbf{74.32}&0.0014	&\textbf{65.89}&\textbf{78.46}&0.0010\\
    \bottomrule
  \end{tabular}
  }
  \end{center}
\end{table*}

\begin{table*}[]
\begin{center}
  \caption{\textcolor{black}{Emotion recognition performance comparison with different network configurations using leave-one-subject-out cross-validation subject-independent protocol on MAHNOB-HCI database.}}
  \label{tab:diffNetworkHCI}
  \scalebox{0.9}{
  \color{black}\begin{tabular}{lcccccccccccc}
    \toprule
           \ \multirow{2}{*}{Networks} & \multicolumn{3}{c}{Valence} & \multicolumn{3}{c}{Arousal} & \multicolumn{3}{c}{Dominance}   & \multicolumn{3}{c}{Predictability} \\
      \  & $P_{acc}$ & $P_f$ &NMI & $P_{acc}$ & $P_f$  &NMI& $P_{acc}$ & $P_f$ &NMI & $P_{acc}$ & $P_f$ &NMI \\
    \midrule
    \ CNN based &58.22&63.99&0.0565&55.88&63.07&0.0462&62.91&74.94&0.0829&71.94&82.28 &0.0867\\
    \ CNN-GAN based &59.37&63.36&0.0545&59.24&61.08&0.0531&63.52&74.83&0.0678&72.41&82.41&0.1028\\
    \ CNN-RNN based &60.08&71.63&0.1087&54.37&63.24&0.0921&66.35&75.94&0.1427&74.35&83.35&0.1687\\
     \ \textbf{CNN-RNN-GAN based (EEGFuseNet)} & \textbf{60.64}	&\textbf{72.18}&\textbf{0.1129} &\textbf{62.06}&62.05&0.0918	&\textbf{67.08}&\textbf{76.65}&\textbf{0.1783}	&\textbf{74.63}&\textbf{83.61}&\textbf{0.1829}\\
    \bottomrule
  \end{tabular}
  }
  \end{center}
\end{table*}

\begin{table}[]
\begin{center}
  \caption{\textcolor{black}{Emotion recognition performance comparison with different network configurations using leave-one-subject-out cross-validation subject-independent protocol on SEED database.}}
  \label{tab:diffNetworkSEED}
  \scalebox{0.8}{
  \color{black}\begin{tabular}{lccccc}
    \toprule
    \  \multirow{2}{*}{Methods} & \multicolumn{3}{c}{Two-Class} & \multicolumn{2}{c}{Three-Class} \\
    \ & $P_{acc}$ & $P_f$ &  NMI & $P_{acc}$ & NMI \\
    \midrule
     \ CNN based & 70.26 & 76.93 & 0.2891 & 52.04 & 0.2260\\
    \ CNN-GAN based  & 76.39 & 79.22 & 0.3528& 53.78 & 0.2969\\
     \ CNN-RNN based  & 74.27 &  76.96 & 0.3229 & 53.76 & 0.2987\\
    %   \ \textbf{CNN-RNN-GAN based}  & \multirow{2}{*}{\textbf{80.83}} &  \multirow{2}{*}{\textbf{82.03}} & \multirow{2}{*}{\textbf{0.4381}} & \multirow{2}{*}{\textbf{59.06}}	& \multirow{2}{*}{\textbf{0.3569}}\\
    %   \ \textbf{(EEGFuseNet)} &&&&&\\
     \ \textbf{CNN-RNN-GAN based (EEGFuseNet)}  & {\textbf{80.83}} &  {\textbf{82.03}} & {\textbf{0.4381}} & {\textbf{59.06}}	& {\textbf{0.3569}}\\
    \bottomrule
  \end{tabular}
  }
  \end{center}
\end{table}

% \begin{table*}[]
% \begin{center}
%   \caption{\textcolor{black}{Emotion recognition performance comparison with different network configurations using leave-one-subject-out cross-validation subject-independent protocol on SEED database.}}
%   \label{tab:diffNetworkSEED}
%   \scalebox{1}{
%   \color{black}\begin{tabular}{lccccc}
%     \toprule
%     \  \multirow{2}{*}{Methods} & \multicolumn{3}{c}{Two-Class} & \multicolumn{2}{c}{Three-Class} \\
%     \ & $P_{acc}$ & $P_f$ &  NMI & $P_{acc}$ & NMI \\
%     \midrule
%      \ CNN based & 70.26 & 76.93 & 0.2891 & 52.04 & 0.2260\\
%     \ CNN-GAN based  & 76.39 & 79.22 & 0.3528& 53.78 & 0.2969\\
%      \ CNN-RNN based  & 74.27 &  76.96 & 0.3229 & 53.76 & 0.2987\\
%       \ CNN-RNN-GAN based (EEGFuseNet)  & \textbf{80.83} &  \textbf{82.03} & \textbf{0.4381} & \textbf{59.06}	&\textbf{0.3569}\\
%     \bottomrule
%   \end{tabular}
%   }
%   \end{center}
% \end{table*}

\color{black}
\subsection{Hyperparameter Effect}
\label{parameterEffect}
\begin{figure}
\begin{center}
\includegraphics[width=0.5\textwidth]{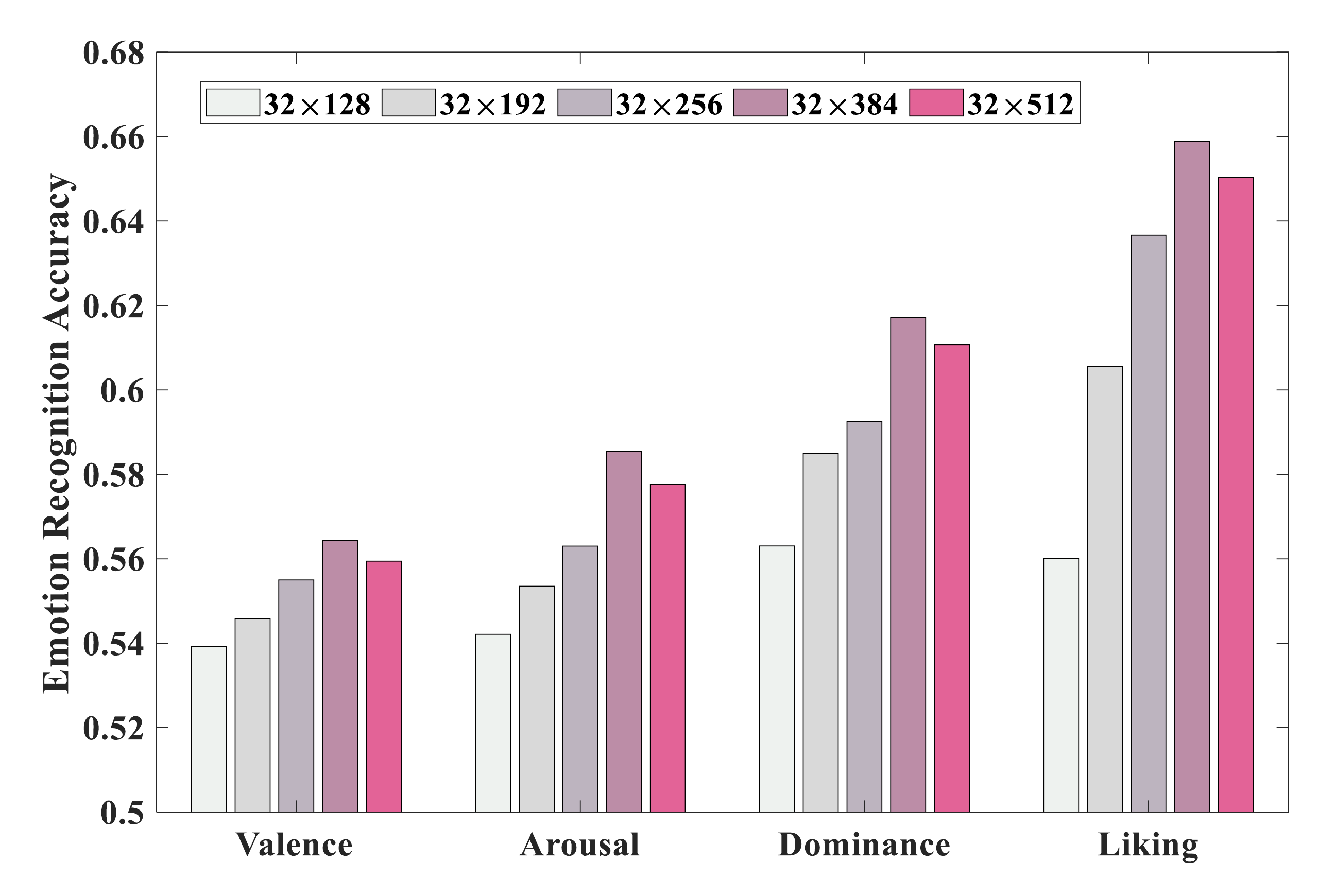}
\end{center}
\caption{\textcolor{black}{A comparison of emotion recognition performance with adjusted input sizes to the proposed EEGFuseNet using leave-one-subject-out cross-validation subject-independent protocol on DEAP database.}}
\label{fig:inputSizeCom}
\end{figure}

\begin{figure}
\begin{center}
\subfloat[]{\includegraphics[width=0.5\textwidth]{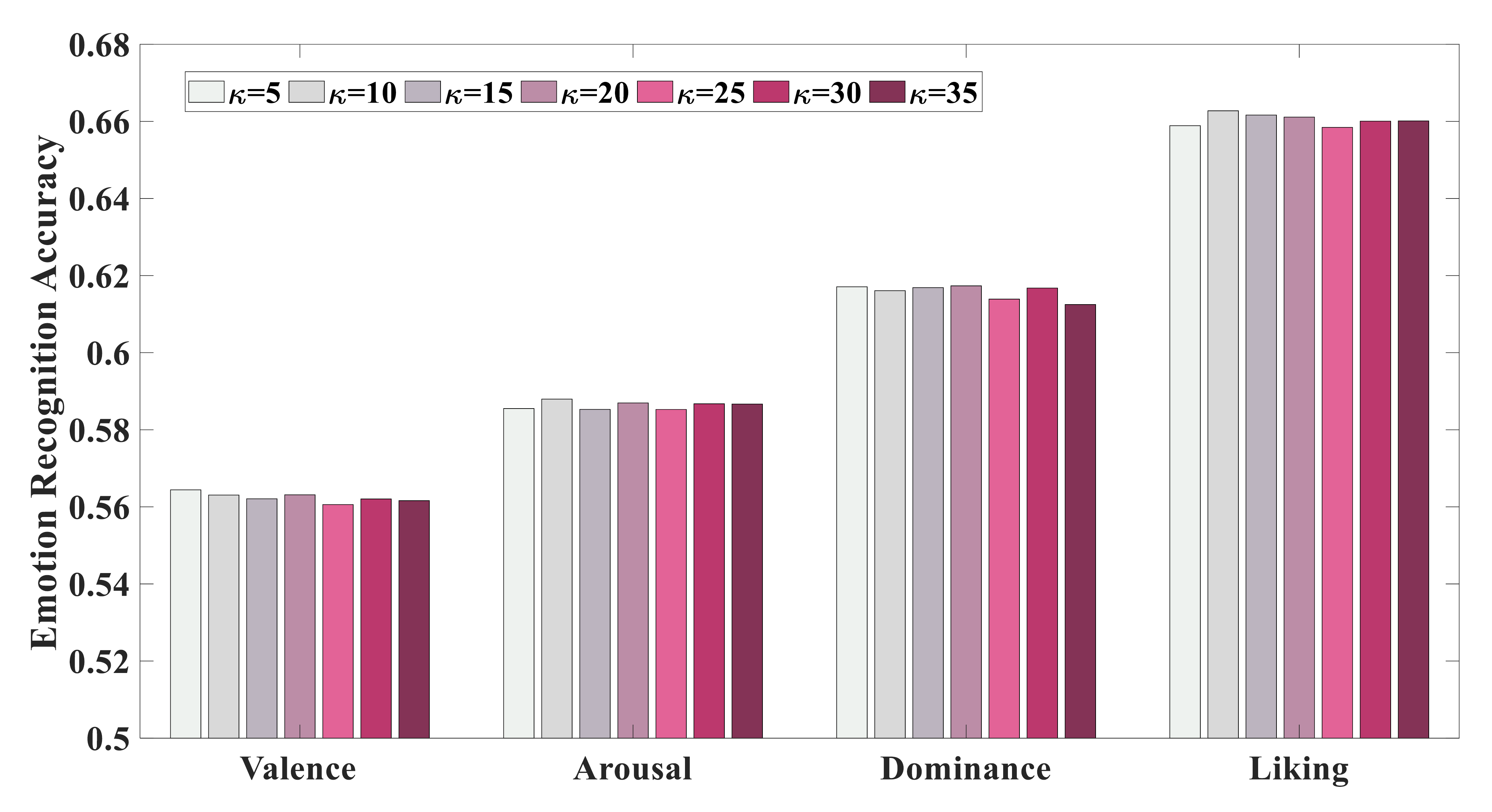}}

\subfloat[]{\includegraphics[width=0.5\textwidth]{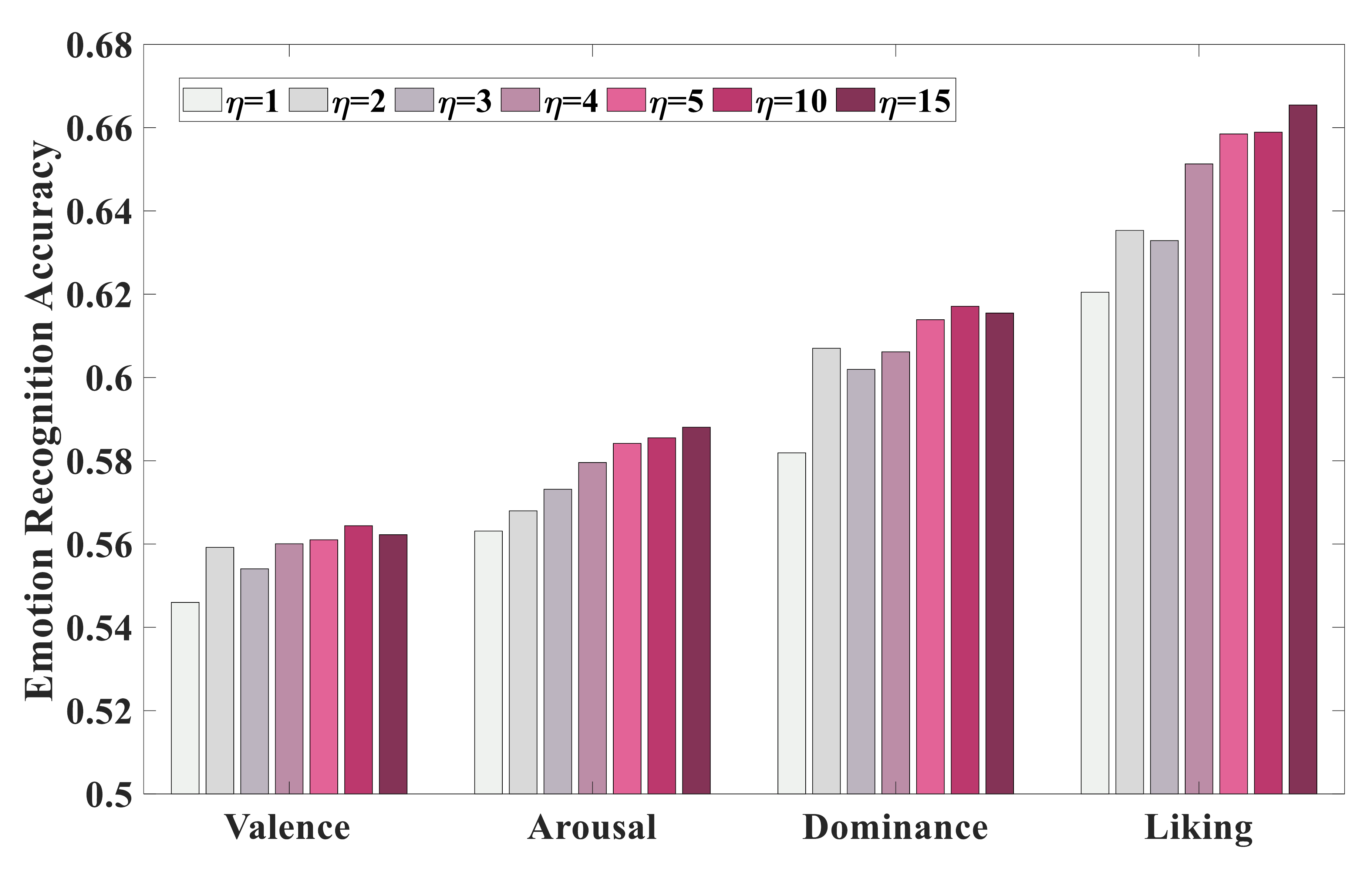}}
\end{center}
\caption{\textcolor{black}{A comparison of emotion recognition performance with various (a) $\kappa$ and (b) $\eta$ values using leave-one-subject-out cross-validation subject-independent protocol on DEAP database.}}
\label{fig:para_var}
\end{figure}

The effect of the hyperparameters in the proposed EEGFuseNet based hypergraph decoding framework on the emotion recognition performance is also validated. For EEGFuseNet, we test input size effect on model performance. Take DEAP database as an example. We adjust the input data size of EEGFuseNet from 32$\times$128 to 32$\times$512 and summarize the corresponding emotion recognition performance in Fig. \ref{fig:inputSizeCom}. According to the network design, the kernel size in the first convolution layer and the depthwise separable convolution layer would be adaptively adjusted according to the input data size. The results show a relative smaller input size of 32$\times$384 perform the best, which is evident to cover almost of the important information in the collected data and improve the computation efficiency as well. If the input data size is further reduced, a loss of information would lead to a significant decrease in the recognition performance. For the original data size (32$\times$512), the corresponding kernel size and the number of parameters in the network greatly increase compared to the other input data sizes, which would also lead to a higher chance of overfitting especially when the sample size is not large enough. \textcolor{black}{Furthermore, we verify the corresponding computational time under different input sizes and report the results in Table \ref{tab:computeTime}. Here, the computational time is separately measured under EEGFuseNet training, EEGFuseNet testing, and hypergraph decoding. The overall computational time is also given. It is found that EEGFuseNet training time is not much affected by the input size, but the computational time of EEGFuseNet testing and hypergraph decoding increase along with an increase of the input size. Note that after model learning, the computational time reduces to seconds level (11.36s $ \sim $ 44.51s) which would be acceptable in real applications.} On the other hand, for hypergraph decoding model, the effect of hyperedge size ($\kappa$) on emotion recognition performance is also examined. We adjust $\kappa$ value from 5 to 35 with a step of 5 and present the corresponding emotion recognition performance in Fig. \ref{fig:para_var} (a). The results show the performance is relative stable and less sensitive to the change of $\kappa$ value.

\begin{table}[]
\begin{center}
  \caption{\textcolor{black}{The computational time (in seconds) with an adjusted input size using leave-one-subject-out cross-validation subject-independent protocol on DEAP database ($\eta=10$).}}
  \label{tab:computeTime}
  \scalebox{0.9}{
  \color{black}\begin{tabular}{ccccc}
    \toprule
      \ \multirow{2}{*}{Input Size} & \multirow{2}{*}{Overall(s)} & EEGFuseNet  & EEGFuseNet  & Hypergraph  \\
      \ & & Training(s) & Testing(s) & Decoding(s)\\
    \midrule
    \ 32 $\times$ 128 &28697.33 &18059.97&11.36	&10626\\
    \ 32 $\times$ 192 &29136.33&18054.39&16.94&11065\\
    \ 32 $\times$ 256 &29292.88&18060.51&24.37&11208\\
     \ 32 $\times$ 384 &29836.99&18084.71&34.28&11718\\
      \ 32 $\times$ 512 &31447.48&18081.97&44.51&13321\\
    \bottomrule
  \end{tabular}
  }
  \end{center}
\end{table}

% \begin{table*}[]
% \begin{center}
%   \caption{\textcolor{black}{The computational time with an adjusted input size using leave-one-subject-out cross-validation subject-independent protocol on DEAP database ($\eta=10$).}}
%   \label{tab:computeTime}
%   \scalebox{1}{
%   \color{black}\begin{tabular}{ccccc}
%     \toprule
%       \ \multirow{2}{*}{Input Size} & \multirow{2}{*}{Overall} & EEGFuseNet  & EEGFuseNet  & Hypergraph  \\
%       \ & & Training & Testing & Decoding\\
%     \midrule
%     \ 32 $\times$ 128 &28697.33 &18059.97&11.36	&10626\\
%     \ 32 $\times$ 192 &29136.33&18054.39&16.94&11065\\
%     \ 32 $\times$ 256 &29292.88&18060.51&24.37&11208\\
%      \ 32 $\times$ 384 &29836.99&18084.71&34.28&11718\\
%       \ 32 $\times$ 512 &31447.48&18081.97&44.51&13321\\
%     \bottomrule
%   \end{tabular}
%   }
%   \end{center}
% \end{table*}

\begin{figure}
\begin{center}
\includegraphics[width=0.4\textwidth]{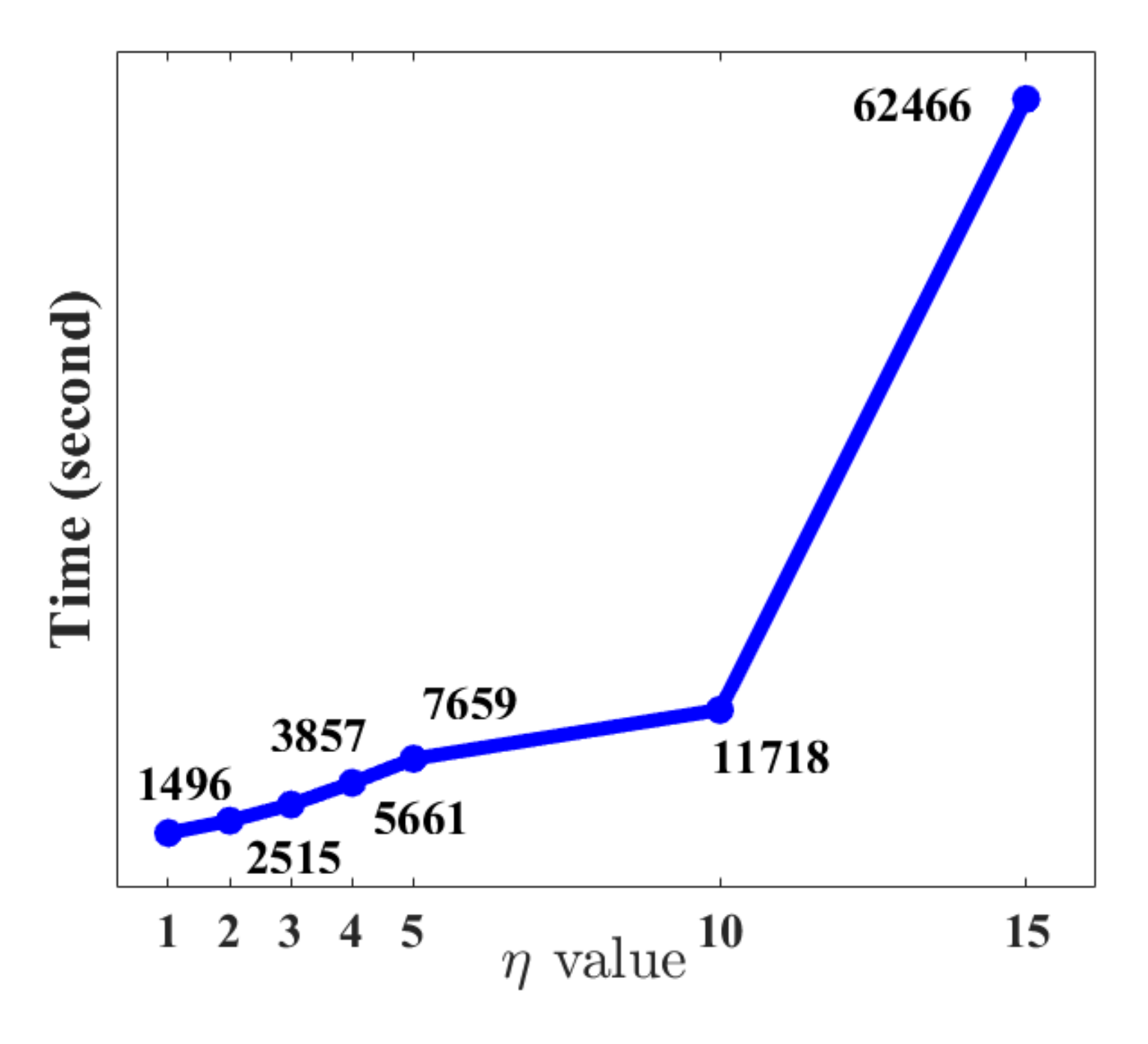}
\end{center}
\caption{\textcolor{black}{The corresponding computational time (in seconds) when $\eta$ value varies from 1 to 15 using leave-one-subject-out cross-validation subject-independent protocol on DEAP database.}}
\label{fig:etaTime}
\end{figure}

\textcolor{black}{As present in Section \ref{sec:experiment}-C, we introduce a speed up theorem with a hyperparameter $\eta \%$ to reduce the computation complexity of hypergraph decoding. To evaluate the effect of $\eta$ on the model learning performance, we adjust the value to 1, 2, 3, 4, 5, 10, and 15, where the corresponding training data size is 744, 1488, 2232, 2976, 3720, 7440 and 11160 samples (the total training data candidates are 74400 samples). The emotion recognition performance under different $\eta$ values are shown in Fig. \ref{fig:para_var} (b). It reveals that an increase of $\eta$ value could generally lead to a greater emotion recognition accuracy. For the case of $\eta=2$ achieving better performance than $\eta=3$, it could be the randomly selected training data of $\eta=2$ probably share similar patterns to the test data and less individual difference are involved. The average computational time under different $\eta$ values is shown in Fig. \ref{fig:etaTime}. An increase of $\eta$ value leads to an exponential growth in the computational time, where the cost time of hypergraph decoding is 1496s for $\eta=1$ and 62466s for $\eta=15$. For the $\eta$ value of 10, the computational time is 11718s. There is a trade-off between decoding performance and computational time.} 

% \begin{figure}
% \begin{center}
% \includegraphics[width=0.5\textwidth]{./hyperedgeSizeResults}
% \end{center}
% \caption{A comparison of emotion recognition performance with various hyperedge size values.}
% \label{fig:hyperedgeSizeCOM}
% \end{figure}

% \begin{figure}
% \begin{center}
% \includegraphics[width=0.5\textwidth]{./sampleRatePerformance}
% \end{center}
% \caption{A comparison of emotion recognition performance with different sampling rates.}
% \label{fig:sampleRateCOM}
% \end{figure}

\color{black}
\subsection{Conclusion}
The aim of this paper is to present a theoretical and practical method for valid and reliable feature characterization and fusion of high-dimensional EEG signals in an unsupervised manner. This paper offers a comprehensive and dedicated comparisons on the proposed EEGFuseNet with different specific designs and configurations. The efficiency and effectivity of the extracted features is demonstrated in an emotion recognition application. The results reveal that the proposed hybrid EEGFuseNet (CNN-RNN-GAN based) systematically outperforms the other networks (CNN based, hybrid CNN-GAN, and hybrid CNN-RNN), which also proves our original hypothesis in the network design. Notably, the proposed characterization, fusion and classification framework is a self-learning paradigm, without any requirement on labelling information in the training process. This work could serve as a foundational framework for high-dimensional EEG study and assess the validity of other unsupervised methods beyond non-stationary time-series EEG signals. \textcolor{black}{On the other hand, due to the lack of label guidance, the performance of current unsupervised results is still lower than those of the supervised methods. There is still a need to further develop unsupervised algorithms for EEG based emotion decoding applications and to enhance the unsupervised performance. Especially for most real-world applications, we don’t have the labels for data or we don’t have a power platform in portable devices to support model re-training when new data are coming. Thus, unsupervised decoding methods would play a significant role for model learning, updating and working.}

\section{Conflicts of Interest}
The authors declare that they have no conflicts of interest.

\section{Acknowledgments}
This study was supported by National Natural Science Foundation of China (No.61906122 and No.91859122).

% \appendices

% % you can choose not to have a title for an appendix
% % if you want by leaving the argument blank
% \section{}


\ifCLASSOPTIONcaptionsoff
  \newpage
\fi

\bibliographystyle{IEEEtran}

\begin{thebibliography}{10}
\providecommand{\url}[1]{#1}
\csname url@samestyle\endcsname
\providecommand{\newblock}{\relax}
\providecommand{\bibinfo}[2]{#2}
\providecommand{\BIBentrySTDinterwordspacing}{\spaceskip=0pt\relax}
\providecommand{\BIBentryALTinterwordstretchfactor}{4}
\providecommand{\BIBentryALTinterwordspacing}{\spaceskip=\fontdimen2\font plus
\BIBentryALTinterwordstretchfactor\fontdimen3\font minus
  \fontdimen4\font\relax}
\providecommand{\BIBforeignlanguage}[2]{{%
\expandafter\ifx\csname l@#1\endcsname\relax
\typeout{** WARNING: IEEEtran.bst: No hyphenation pattern has been}%
\typeout{** loaded for the language `#1'. Using the pattern for}%
\typeout{** the default language instead.}%
\else
\language=\csname l@#1\endcsname
\fi
#2}}
\providecommand{\BIBdecl}{\relax}
\BIBdecl

\bibitem{alarcao2017emotions}
S.~M. Alarcao and M.~J. Fonseca, ``Emotions recognition using eeg signals: A
  survey,'' \emph{IEEE Transactions on Affective Computing}, vol.~10, no.~3,
  pp. 374--393, 2017.

\bibitem{hu2020video}
W.~Hu, G.~Huang, L.~Li, L.~Zhang, Z.~Zhang, and Z.~Liang, ``Video-triggered
  eeg-emotion public databases and current methods: A survey,'' \emph{Brain
  Science Advances}, vol.~6, no.~3, pp. 255--287, 2020.

\bibitem{haufe2013critical}
S.~Haufe, V.~V. Nikulin, K.-R. M{\"u}ller, and G.~Nolte, ``A critical
  assessment of connectivity measures for eeg data: a simulation study,''
  \emph{Neuroimage}, vol.~64, pp. 120--133, 2013.

\bibitem{milz2016functional}
P.~Milz, P.~L. Faber, D.~Lehmann, T.~Koenig, K.~Kochi, and R.~D.
  Pascual-Marqui, ``The functional significance of eeg
  microstates—associations with modalities of thinking,'' \emph{Neuroimage},
  vol. 125, pp. 643--656, 2016.

\bibitem{castelnovo2016scalp}
A.~Castelnovo, B.~A. Riedner, R.~F. Smith, G.~Tononi, M.~Boly, and R.~M. Benca,
  ``Scalp and source power topography in sleepwalking and sleep terrors: a
  high-density eeg study,'' \emph{Sleep}, vol.~39, no.~10, pp. 1815--1825,
  2016.

\bibitem{ma2017eeg}
X.~Ma, X.~Huang, Y.~Shen, Z.~Qin, Y.~Ge, Y.~Chen, and X.~Ning, ``Eeg based
  topography analysis in string recognition task,'' \emph{Physica A:
  Statistical Mechanics and its Applications}, vol. 469, pp. 531--539, 2017.

\bibitem{ramos2020feature}
R.~Ramos-Aguilar, J.~A. Olvera-L{\'o}pez, I.~Olmos-Pineda, and
  S.~S{\'a}nchez-Urrieta, ``Feature extraction from eeg spectrograms for
  epileptic seizure detection,'' \emph{Pattern Recognition Letters}, vol. 133,
  pp. 202--209, 2020.

\bibitem{islam2019wavelet}
M.~R. Islam and M.~Ahmad, ``Wavelet analysis based classification of emotion
  from eeg signal,'' in \emph{2019 International Conference on Electrical,
  Computer and Communication Engineering (ECCE)}.\hskip 1em plus 0.5em minus
  0.4em\relax IEEE, 2019, pp. 1--6.

\bibitem{jirayucharoensak2014eeg}
S.~Jirayucharoensak, S.~Pan-Ngum, and P.~Israsena, ``Eeg-based emotion
  recognition using deep learning network with principal component based
  covariate shift adaptation,'' \emph{The Scientific World Journal}, vol. 2014,
  2014.

\bibitem{zheng2015investigating}
W.-L. Zheng and B.-L. Lu, ``Investigating critical frequency bands and channels
  for eeg-based emotion recognition with deep neural networks,'' \emph{IEEE
  Transactions on Autonomous Mental Development}, vol.~7, no.~3, pp. 162--175,
  2015.

\bibitem{schirrmeister2017deep}
R.~T. Schirrmeister, J.~T. Springenberg, L.~D.~J. Fiederer, M.~Glasstetter,
  K.~Eggensperger, M.~Tangermann, F.~Hutter, W.~Burgard, and T.~Ball, ``Deep
  learning with convolutional neural networks for eeg decoding and
  visualization,'' \emph{Human brain mapping}, vol.~38, no.~11, pp. 5391--5420,
  2017.

\bibitem{song2018eeg}
T.~Song, W.~Zheng, P.~Song, and Z.~Cui, ``Eeg emotion recognition using
  dynamical graph convolutional neural networks,'' \emph{IEEE Transactions on
  Affective Computing}, vol.~11, no.~3, pp. 532--541, 2018.

\bibitem{cimtay2020investigating}
Y.~Cimtay and E.~Ekmekcioglu, ``Investigating the use of pretrained
  convolutional neural network on cross-subject and cross-dataset eeg emotion
  recognition,'' \emph{Sensors}, vol.~20, no.~7, p. 2034, 2020.

\bibitem{luo2017unsupervised}
Z.~Luo, B.~Peng, D.-A. Huang, A.~Alahi, and L.~Fei-Fei, ``Unsupervised learning
  of long-term motion dynamics for videos,'' in \emph{Proceedings of the IEEE
  conference on computer vision and pattern recognition}, 2017, pp. 2203--2212.

\bibitem{barlow1989unsupervised}
H.~B. Barlow, ``Unsupervised learning,'' \emph{Neural computation}, vol.~1,
  no.~3, pp. 295--311, 1989.

\bibitem{liang2019unsupervised}
Z.~Liang, S.~Oba, and S.~Ishii, ``An unsupervised eeg decoding system for human
  emotion recognition,'' \emph{Neural Networks}, vol. 116, pp. 257--268, 2019.

\bibitem{tao2015unsupervised}
C.~Tao, H.~Pan, Y.~Li, and Z.~Zou, ``Unsupervised spectral--spatial feature
  learning with stacked sparse autoencoder for hyperspectral imagery
  classification,'' \emph{IEEE Geoscience and remote sensing letters}, vol.~12,
  no.~12, pp. 2438--2442, 2015.

\bibitem{kiran2018overview}
B.~R. Kiran, D.~M. Thomas, and R.~Parakkal, ``An overview of deep learning
  based methods for unsupervised and semi-supervised anomaly detection in
  videos,'' \emph{Journal of Imaging}, vol.~4, no.~2, p.~36, 2018.

\bibitem{deng2014autoencoder}
J.~Deng, Z.~Zhang, F.~Eyben, and B.~Schuller, ``Autoencoder-based unsupervised
  domain adaptation for speech emotion recognition,'' \emph{IEEE Signal
  Processing Letters}, vol.~21, no.~9, pp. 1068--1072, 2014.

\bibitem{wen2018deep}
T.~Wen and Z.~Zhang, ``Deep convolution neural network and autoencoders-based
  unsupervised feature learning of eeg signals,'' \emph{IEEE Access}, vol.~6,
  pp. 25\,399--25\,410, 2018.

\bibitem{shoeibi2021comprehensive}
A.~Shoeibi, N.~Ghassemi, R.~Alizadehsani, M.~Rouhani, H.~Hosseini-Nejad,
  A.~Khosravi, M.~Panahiazar, and S.~Nahavandi, ``A comprehensive comparison of
  handcrafted features and convolutional autoencoders for epileptic seizures
  detection in eeg signals,'' \emph{Expert Systems with Applications}, vol.
  163, p. 113788, 2021.

\bibitem{tabar2016novel}
Y.~R. Tabar and U.~Halici, ``A novel deep learning approach for classification
  of eeg motor imagery signals,'' \emph{Journal of neural engineering},
  vol.~14, no.~1, p. 016003, 2016.

\bibitem{ye2019understanding}
J.~C. Ye and W.~K. Sung, ``Understanding geometry of encoder-decoder cnns,'' in
  \emph{International Conference on Machine Learning}.\hskip 1em plus 0.5em
  minus 0.4em\relax PMLR, 2019, pp. 7064--7073.

\bibitem{jiao2018deep}
Z.~Jiao, X.~Gao, Y.~Wang, J.~Li, and H.~Xu, ``Deep convolutional neural
  networks for mental load classification based on eeg data,'' \emph{Pattern
  Recognition}, vol.~76, pp. 582--595, 2018.

\bibitem{chen2019deep}
H.~Chen, Y.~Song, and X.~Li, ``A deep learning framework for identifying
  children with adhd using an eeg-based brain network,'' \emph{Neurocomputing},
  vol. 356, pp. 83--96, 2019.

\bibitem{lawhern2018eegnet}
V.~J. Lawhern, A.~J. Solon, N.~R. Waytowich, S.~M. Gordon, C.~P. Hung, and
  B.~J. Lance, ``Eegnet: a compact convolutional neural network for eeg-based
  brain--computer interfaces,'' \emph{Journal of neural engineering}, vol.~15,
  no.~5, p. 056013, 2018.

\bibitem{rifai2011contractive}
S.~Rifai, P.~Vincent, X.~Muller, X.~Glorot, and Y.~Bengio, ``Contractive
  auto-encoders: Explicit invariance during feature extraction,'' in
  \emph{Icml}, 2011.

\bibitem{gehring2013extracting}
J.~Gehring, Y.~Miao, F.~Metze, and A.~Waibel, ``Extracting deep bottleneck
  features using stacked auto-encoders,'' in \emph{2013 IEEE international
  conference on acoustics, speech and signal processing}.\hskip 1em plus 0.5em
  minus 0.4em\relax IEEE, 2013, pp. 3377--3381.

\bibitem{akbari2018semi}
M.~Akbari and J.~Liang, ``Semi-recurrent cnn-based vae-gan for sequential data
  generation,'' in \emph{2018 IEEE International Conference on Acoustics,
  Speech and Signal Processing (ICASSP)}.\hskip 1em plus 0.5em minus
  0.4em\relax IEEE, 2018, pp. 2321--2325.

\bibitem{makhzani2015adversarial}
A.~Makhzani, J.~Shlens, N.~Jaitly, I.~Goodfellow, and B.~Frey, ``Adversarial
  autoencoders,'' \emph{arXiv preprint arXiv:1511.05644}, 2015.

\bibitem{chen2018unsupervised}
X.~Chen and E.~Konukoglu, ``Unsupervised detection of lesions in brain mri
  using constrained adversarial auto-encoders,'' \emph{arXiv preprint
  arXiv:1806.04972}, 2018.

\bibitem{sahu2018adversarial}
S.~Sahu, R.~Gupta, G.~Sivaraman, W.~AbdAlmageed, and C.~Espy-Wilson,
  ``Adversarial auto-encoders for speech based emotion recognition,''
  \emph{arXiv preprint arXiv:1806.02146}, 2018.

\bibitem{serban2017hierarchical}
I.~Serban, A.~Sordoni, R.~Lowe, L.~Charlin, J.~Pineau, A.~Courville, and
  Y.~Bengio, ``A hierarchical latent variable encoder-decoder model for
  generating dialogues,'' in \emph{Proceedings of the AAAI Conference on
  Artificial Intelligence}, vol.~31, no.~1, 2017.

\bibitem{vosoughi2016tweet2vec}
S.~Vosoughi, P.~Vijayaraghavan, and D.~Roy, ``Tweet2vec: Learning tweet
  embeddings using character-level cnn-lstm encoder-decoder,'' in
  \emph{Proceedings of the 39th International ACM SIGIR conference on Research
  and Development in Information Retrieval}, 2016, pp. 1041--1044.

\bibitem{li2017deep}
P.~Li, W.~Lam, L.~Bing, and Z.~Wang, ``Deep recurrent generative decoder for
  abstractive text summarization,'' \emph{arXiv preprint arXiv:1708.00625},
  2017.

\bibitem{cho2014properties}
K.~Cho, B.~Van~Merri{\"e}nboer, D.~Bahdanau, and Y.~Bengio, ``On the properties
  of neural machine translation: Encoder-decoder approaches,'' \emph{arXiv
  preprint arXiv:1409.1259}, 2014.

\bibitem{chung2014empirical}
J.~Chung, C.~Gulcehre, K.~Cho, and Y.~Bengio, ``Empirical evaluation of gated
  recurrent neural networks on sequence modeling,'' \emph{arXiv preprint
  arXiv:1412.3555}, 2014.

\bibitem{zhou2006learning}
D.~Zhou, J.~Huang, and B.~Sch{\"o}lkopf, ``Learning with hypergraphs:
  Clustering, classification, and embedding,'' \emph{Advances in neural
  information processing systems}, vol.~19, pp. 1601--1608, 2006.

\bibitem{koelstra2011deap}
S.~Koelstra, C.~Muhl, M.~Soleymani, J.-S. Lee, A.~Yazdani, T.~Ebrahimi, T.~Pun,
  A.~Nijholt, and I.~Patras, ``Deap: A database for emotion analysis; using
  physiological signals,'' \emph{IEEE transactions on affective computing},
  vol.~3, no.~1, pp. 18--31, 2011.

\bibitem{soleymani2011multimodal}
M.~Soleymani, J.~Lichtenauer, T.~Pun, and M.~Pantic, ``A multimodal database
  for affect recognition and implicit tagging,'' \emph{IEEE transactions on
  affective computing}, vol.~3, no.~1, pp. 42--55, 2011.

\bibitem{glorot2010understanding}
X.~Glorot and Y.~Bengio, ``Understanding the difficulty of training deep
  feedforward neural networks,'' in \emph{Proceedings of the thirteenth
  international conference on artificial intelligence and statistics}.\hskip
  1em plus 0.5em minus 0.4em\relax JMLR Workshop and Conference Proceedings,
  2010, pp. 249--256.

\bibitem{liu2012eeg}
Y.~Liu and O.~Sourina, ``Eeg-based valence level recognition for real-time
  applications,'' in \emph{2012 International Conference on Cyberworlds}.\hskip
  1em plus 0.5em minus 0.4em\relax IEEE, 2012, pp. 53--60.

\bibitem{li2015eeg}
X.~Li, P.~Zhang, D.~Song, G.~Yu, Y.~Hou, and B.~Hu, ``Eeg based emotion
  identification using unsupervised deep feature learning,'' 2015.

\bibitem{chen2015electroencephalogram}
J.~Chen, B.~Hu, P.~Moore, X.~Zhang, and X.~Ma, ``Electroencephalogram-based
  emotion assessment system using ontology and data mining techniques,''
  \emph{Applied Soft Computing}, vol.~30, pp. 663--674, 2015.

\bibitem{bahari2013eeg}
F.~Bahari and A.~Janghorbani, ``Eeg-based emotion recognition using recurrence
  plot analysis and k nearest neighbor classifier,'' in \emph{2013 20th Iranian
  Conference on Biomedical Engineering (ICBME)}.\hskip 1em plus 0.5em minus
  0.4em\relax IEEE, 2013, pp. 228--233.

\bibitem{naser2013recognition}
D.~S. Naser and G.~Saha, ``Recognition of emotions induced by music videos
  using dt-cwpt,'' in \emph{2013 Indian Conference on Medical Informatics and
  Telemedicine (ICMIT)}.\hskip 1em plus 0.5em minus 0.4em\relax IEEE, 2013, pp.
  53--57.

\bibitem{zhuang2017emotion}
N.~Zhuang, Y.~Zeng, L.~Tong, C.~Zhang, H.~Zhang, and B.~Yan, ``Emotion
  recognition from eeg signals using multidimensional information in emd
  domain,'' \emph{BioMed research international}, vol. 2017, 2017.

\bibitem{torres2014comparative}
C.~A. Torres-Valencia, H.~F. Garcia-Arias, M.~A.~A. Lopez, and A.~A.
  Orozco-Guti{\'e}rrez, ``Comparative analysis of physiological signals and
  electroencephalogram (eeg) for multimodal emotion recognition using
  generative models,'' in \emph{2014 XIX Symposium on Image, Signal Processing
  and Artificial Vision}.\hskip 1em plus 0.5em minus 0.4em\relax IEEE, 2014,
  pp. 1--5.

\bibitem{atkinson2016improving}
J.~Atkinson and D.~Campos, ``Improving bci-based emotion recognition by
  combining eeg feature selection and kernel classifiers,'' \emph{Expert
  Systems with Applications}, vol.~47, pp. 35--41, 2016.

\bibitem{liu2016emotion}
J.~Liu, H.~Meng, A.~Nandi, and M.~Li, ``Emotion detection from eeg
  recordings,'' in \emph{2016 12th International Conference on Natural
  Computation, Fuzzy Systems and Knowledge Discovery (ICNC-FSKD)}.\hskip 1em
  plus 0.5em minus 0.4em\relax IEEE, 2016, pp. 1722--1727.

\bibitem{shahnaz2016emotion}
C.~Shahnaz, S.~S. Hasan \emph{et~al.}, ``Emotion recognition based on wavelet
  analysis of empirical mode decomposed eeg signals responsive to music
  videos,'' in \emph{2016 IEEE Region 10 Conference (TENCON)}.\hskip 1em plus
  0.5em minus 0.4em\relax IEEE, 2016, pp. 424--427.

\bibitem{chen2019hierarchical}
J.~Chen, D.~Jiang, and Y.~Zhang, ``A hierarchical bidirectional gru model with
  attention for eeg-based emotion classification,'' \emph{IEEE Access}, vol.~7,
  pp. 118\,530--118\,540, 2019.

\bibitem{zhong2020cross}
X.~Zhong, Z.~Yin, and J.~Zhang, ``Cross-subject emotion recognition from eeg
  using convolutional neural networks,'' in \emph{2020 39th Chinese Control
  Conference (CCC)}.\hskip 1em plus 0.5em minus 0.4em\relax IEEE, 2020, pp.
  7516--7521.

\bibitem{du2020efficient}
X.~Du, C.~Ma, G.~Zhang, J.~Li, Y.-K. Lai, G.~Zhao, X.~Deng, Y.-J. Liu, and
  H.~Wang, ``An efficient lstm network for emotion recognition from
  multichannel eeg signals,'' \emph{IEEE Transactions on Affective Computing},
  2020.

\bibitem{zhu2014emotion}
Y.~Zhu, S.~Wang, and Q.~Ji, ``Emotion recognition from users' eeg signals with
  the help of stimulus videos,'' in \emph{2014 IEEE international conference on
  multimedia and expo (ICME)}.\hskip 1em plus 0.5em minus 0.4em\relax IEEE,
  2014, pp. 1--6.

\bibitem{huang2016multi}
X.~Huang, J.~Kortelainen, G.~Zhao, X.~Li, A.~Moilanen, T.~Sepp{\"a}nen, and
  M.~Pietik{\"a}inen, ``Multi-modal emotion analysis from facial expressions
  and electroencephalogram,'' \emph{Computer Vision and Image Understanding},
  vol. 147, pp. 114--124, 2016.

\bibitem{yin2020locally}
Z.~Yin, L.~Liu, J.~Chen, B.~Zhao, and Y.~Wang, ``Locally robust eeg feature
  selection for individual-independent emotion recognition,'' \emph{Expert
  Systems with Applications}, vol. 162, p. 113768, 2020.

\bibitem{li2019domain}
J.~Li, S.~Qiu, C.~Du, Y.~Wang, and H.~He, ``Domain adaptation for eeg emotion
  recognition based on latent representation similarity,'' \emph{IEEE
  Transactions on Cognitive and Developmental Systems}, vol.~12, no.~2, pp.
  344--353, 2019.

\bibitem{zheng2016multichannel}
W.~Zheng, ``Multichannel eeg-based emotion recognition via group sparse
  canonical correlation analysis,'' \emph{IEEE Transactions on Cognitive and
  Developmental Systems}, vol.~9, no.~3, pp. 281--290, 2016.

\bibitem{li2018novel}
Y.~Li, W.~Zheng, Z.~Cui, T.~Zhang, and Y.~Zong, ``A novel neural network model
  based on cerebral hemispheric asymmetry for eeg emotion recognition.'' in
  \emph{IJCAI}, 2018, pp. 1561--1567.

\bibitem{li2019regional}
Y.~Li, W.~Zheng, L.~Wang, Y.~Zong, and Z.~Cui, ``From regional to global brain:
  A novel hierarchical spatial-temporal neural network model for eeg emotion
  recognition,'' \emph{IEEE Transactions on Affective Computing}, 2019.

\bibitem{pan2010domain}
S.~J. Pan, I.~W. Tsang, J.~T. Kwok, and Q.~Yang, ``Domain adaptation via
  transfer component analysis,'' \emph{IEEE transactions on neural networks},
  vol.~22, no.~2, pp. 199--210, 2010.

\bibitem{shah2017robust}
S.~A. Shah and V.~Koltun, ``Robust continuous clustering,'' \emph{Proceedings
  of the National Academy of Sciences}, vol. 114, no.~37, pp. 9814--9819, 2017.

\bibitem{zhang2012graph}
W.~Zhang, X.~Wang, D.~Zhao, and X.~Tang, ``Graph degree linkage: Agglomerative
  clustering on a directed graph,'' in \emph{European Conference on Computer
  Vision}.\hskip 1em plus 0.5em minus 0.4em\relax Springer, 2012, pp. 428--441.

\bibitem{yang2016joint}
J.~Yang, D.~Parikh, and D.~Batra, ``Joint unsupervised learning of deep
  representations and image clusters,'' in \emph{Proceedings of the IEEE
  conference on computer vision and pattern recognition}, 2016, pp. 5147--5156.

\end{thebibliography}
\end{document}